\documentclass[
showpacs,
 amsmath,amssymb,
 aps,
 10pt,
twocolumn
]{revtex4-1}

\usepackage[usenames,dvipsnames]{color}
\usepackage{graphicx,textcase} 
\usepackage{dcolumn,overpic} 
\usepackage{bm,color}
\usepackage{chngcntr}
\usepackage[para]{threeparttable}
\usepackage{etoolbox}
\usepackage{lipsum}
\usepackage[bottom]{footmisc}
\usepackage{setspace}
\AtEndEnvironment{table*}{\vskip10pt}{}{}

\newcommand{\pt}[1]{{\color{blue}[PT: #1]}} 

\usepackage{url}
\usepackage{hyperref}

\hypersetup{colorlinks=true,breaklinks,linkcolor=blue,urlcolor=blue,citecolor=blue}

\begin{document}

\title{A systematic study of the local minima in L(S)DA+$U$}
\author{Samara~Keshavarz, Patrik Thunstr\"om}

\affiliation{Uppsala University, Department of Physics and Astronomy, Division of Materials Theory, Box 516, SE-751 20 Uppsala, Sweden}

\date{\today}

\begin{abstract}
We have performed a systematic study of the emergence of meta-stable states in density functional theory plus Hubbard $U$ (DFT+$U$) simulations of NiO, CoO, FeO. Particular attention is given to the spin-polarization of the exchange-correlation functional and the double counting term, and the role of the spin-orbit coupling. The method of occupation matrix control is extended to use constrained random density matrices to map out the local minima in the total energy landscape. The extended scheme, random density matrix control, is successfully benchmarked against UO$_2$, one of the most investigated systems in the field. When applied to the transition metal oxides it yields several meta-stable states which are well-characterized by their local spin and orbital moments. We find that the addition of spin-orbit coupling helps the simulations to converge to the global high-spin energy minimum. The random density matrix control scheme combined with LDA+U yields accurate magnetic moments for all the studied AFM transition metal oxides.
\end{abstract}


\maketitle

\section{Introduction}
Density functional theory augmented with a Hubbard $U$ term (DFT+$U$) has, since its introduction, been successfully applied to a large variety of strongly correlated systems such as transition metal oxides, rare-earth nitrides, and rare-earth oxides~\cite{cmo,pickett-uge,mazin,amadon-ce}. 
This formalism favor an explicit orbital anisotropy according to Hund's rule, which is usually suppressed in the conventional DFT formalism~\cite{ldau1,ldau2,solov,peters}. 
However, the additional degrees of freedom introduced by the orbital polarization is known to cause the emergence of meta-stable states~\cite{coco}. In other words, the total energy landscape may present a multitude local minima at which the self-consistent update of the effective DFT+$U$ potential can get trapped~\cite{mult-lars}. 

The introduction of the additional local Hartree-Fock-like term in the DFT+$U$ Hamiltonian implies a double counting of the local Coulomb interaction. In the case of well-localized orbitals the most popular double counting correction (DC) is the Fully Localized Limit (FLL)~\cite{dc-fll}, which corresponds to a spherical average of the Hartree-Fock interaction under the assumption that the local density matrix is idempotent. One may choose to either allow both the DC term and the exchange-correlation (XC) functional to depend on the charge and magnetization density (sDFT+$U$)\cite{czyzyk1994}, or that both are evaluated using the charge density (cDFT+$U$)\cite{ldau1}. The two choices seek to remove the exchange splitting produced by the local Hartree-Fock term or the XC functional, respectively. The rational for cDFT+$U$ is that it treats spin and orbital moment on equal footing, and that sDFT+$U$ has a tendency to overestimate the local exchange interaction~\cite{lda-lsda,samara-6}.

There are several different mechanism through which spin-orbit coupling (SOC) can affect the abundance of meta-stable states. First, the formation of local minima can be expected to be particularly severe if there is a mismatch between the irreducible representations of the correlated orbitals and their nominal filling due to Hund's rules\pt{cite?}. The inclusion of SOC results in a new set of irreducible representations of lower order, given by the double group, which helps the matching with Hund's rules. Secondly, the coupling between the relative direction of the spin and the orbital moments introduce an additional low-energy degree of freedom, which can cause an increase in the number of meta-stable states. Third, a large SOC may shift the energies of some partially filled local orbitals completely above or below the Fermi energy, and thereby instead reduce the low-energy degrees of freedom. 
It should be noted that orbital polarization in DFT+$U$ does not require SOC, as the local Hartree-Fock potential itself promotes the the filling of the orbitals according to Hund's rules~\cite{solov}. However, in order to find an orbitally polarized state in a scalar relativistic simulation it is often necessary to introduce an orbital asymmetry already in the initial potential.

Several groups have carefully investigated the meta-stable solutions characterizing the sDFT+$U$ approach \cite{amadon-ce,amadon-pt,amadon-uo2,koper-coo,koper-anis,koper-cs,gamma-ce,pickett-uge}, and several methods have been suggested to find the global minimum. For instance, Meredig et al.~\cite{u-ramping} introduced the $U$-ramping method, in which the Hubbard $U$ term is adiabatically turned on to make the fractional orbital occupation gradually converge to integer occupations. However, the ramping procedure requires that the sDFT starting point already captures the correct orbital ordering, as the sDFT+$U$ calculations become difficult to converge when two different orbital orders cross in energy. A second method, proposed by Gryaznov et al.~\cite{cont-sym-red}, is based on a controlled symmetry reduction. In this approach, a structural distortion of the system is initially applied to split the degeneracy of competing meta-stable states. The deformation is then gradually removed. This method is useful to single out known minima in order to perform meaningful comparisons between different structures. The third method is the quasi-annealing method proposed by Geng et al.~\cite{kinetic}. In this approach an auxiliary kinetic energy is added to the system to allow the simulation to escape from the local minima and better explore the energy landscape. The auxiliary kinetic energy is then gradually turned off in the hope that the simulation converges to the global minimum. A fourth method is the occupation matrix control (OMC) scheme~\cite{amadon-pt}. In this approach one seeds the DFT+$U$ potential with a fixed occupation matrix for the correlated orbitals under consideration. After a few iterations, this constraint is lifted and the calculations are left to converge on their own. Depending on the starting occupation matrix, different (meta-stable) states are obtained. 

Despite the large number of studies, a systematic analysis of the energy landscape of sDFT+$U$ and cDFT+$U$, with and without SOC, in $3d$ systems is still lacking. The OMC scheme can potentially capture all the minima in the system, but its systematic application has so far been limited to diagonal initial occupations matrices~\cite{amadon-uo2,u-ramping}. Even when off-diagonal elements were considered, they were selected ad hoc with respect to the material under consideration. While this approach may be adequate for very localized $f$-electrons in highly symmetric crystal structures, it appears to be more questionable for less localized $d$-electrons in less symmetric systems, such as the late transition metal oxides in their experimental crystal structures. For these, a more general procedure to generate the initial occupation matrices is called for, in order to explore the full energy landscape of the system.

The aim of the present study is to devise a general scheme to explore possibly all the minima appearing in sDFT+$U$ and cDFT+$U$, and apply it to the transition metal monoxides (TMO) NiO, CoO, and FeO. To this end, we extend the OMC method to use randomly generated but partially constrained initial density matrices. We benchmark our approach against one of the most investigated systems in the field, UO$_2$~\cite{amadon-uo2,u-ramping}, before we apply it to the TMOs.

This paper is organized as follows. In Sec.~\ref{theory}, we first present the computational details as well as a brief explanation of the theoretical background. This will be followed by the presentation of our approach to seed and monitor the DFT+$U$ simulations to map out the energy landscape. The UO$_2$ benchmark of this extended scheme is presented in Appendix~\ref{app:benchmark}, and the application to NiO, CoO, and FeO is presented in Sec.~\ref{result}. Finally, we discuss the results and draw our conclusion in Sec.~\ref{conclusions}. Appendix~\ref{app:kpoints} shows the importance of a dense $k$-point mesh, exemplified by the trapping of intermediate-spin states in FeO.

\section{Computational Details}
\label{theory}
Our results are based upon the DFT+$U$ scheme implemented in the full-potential linear muffin-tin orbital (FP-LMTO) code RSPt~\cite{rspt-web,rspt-book,rspt-lda+u,rspt-oscar}. The local orbitals are selected as the projections of the LMTO orbitals within a given muffin tin sphere, the so called Muffin Tin Heads (MT)\cite{rspt-lda+u,rspt-oscar}. The local density approximation (LDA) and its spin-dependent counterpart (LSDA) are used as the exchange-correlation functionals for cDFT+$U$ and sDFT+$U$, respectively, together with the FLL double counting correction. The results have been checked using the generalized gradient approximation, with only minor quantitative changes. The LMTO orbitals are generated from a spin-average potential to keep the basis set spin-independent. The k-points in the Brillouin zone are distributed in a Monkhorst-Pack grid, and the integration is performed using Fermi smearing with $T = 70$ K to ensure that the integration respects the crystal symmetries. No explicit symmetry constraints are considered in the TMO calculations in order to allow for symmetry-broken solutions. The TMO and UO$_2$ calculations were performed with very dense $k$-point meshes, 40$\times$40$\times$40 and 16$\times$16$\times$16, respectively. As shown for FeO in Appendix~\ref{app:kpoints}, the use of a more moderate $k$-mesh of 24$\times$24$\times$24 results in a spurious trapping of the simulations in the flatter parts of the energy landscape.

The DFT+$U$ Hamiltonian is given by
\begin{equation}
\label{eqn:h}
H_{DFT+U}= H_{0} + H_{XC} + H_{SOC} + H_U - H_{DC},
\end{equation}
where $H_{0}$ contains the contribution from the kinetic, Hartree, and external potential terms, $H_{SOC}$ is the spin-orbit interaction, $H_{XC}$ is the effective XC potential, $H_U$ is the Hubbard term, and $H_{DC}$ is the FLL DC term. The fully rotationally invariant formulation of DFT+$U$~\cite{4-index-u,ldau2} has been used throughout this paper. The FLL term takes the following form~\cite{dc-fll}
\begin{equation}
\label{dc}
H_{DC}=\frac{1}{2}(U(2N-1)-J(N-1)) - J\frac{\vec{M}\cdot \vec{\sigma}}{2}
\end{equation}
where $N$, $\vec{M}$ and $\vec{\sigma}$ are respectively the total number of electrons in the correlated local orbitals, the corresponding magnetization, and the Pauli matrices. In cDFT+$U$ the magnetization $\vec{M}$ is set to zero when $H_{XC}$ and $H_{DC}$ are evaluated.

We parameterize the $U$-matrix via the Slater integrals $F_0$, $F_2$, $F_4$, and $F_6$, which are in turn constructed from two effective parameters $U$ and $J$, as described in Ref.~\cite{ldau2}. It is common to find a broad range of values used in the literature for the $U$ and $J$ parameters, sometimes in the form of a scan over an effective $U_\textit{eff}$ ($U-J$). In the case of the TMOs $U_\textit{eff}$ ranging from 2 to 8 eV has been used in order to model different properties such as structural, magnetic and electronic properties~\cite{coco,u-tmo1,u-tmo2,ldau1}. In this work we fix the value of $U$ to 6 eV and $J$ to 0.9 eV for the TMOs to simplify the comparison of their local minima. For UO$_2$ we set $U = 4.5$ eV and $J = 0.5$ eV as determined by Kotani et al.~\cite{u-j-uo2} based on an analysis of X-ray photoemission data.

In the TMOs we adopt the experimental anti-ferromagnetic [111] ordering, {\em i.e.} a stacking of alternating ferromagnetic planes along the [111] axis~\cite{feo1}. For the case of UO$_2$, we adopt the simple 1$\mathbf{k}$ [001] anti-ferromagnetic order to be able to compare our results with previous studies~\cite{amadon-uo2,u-ramping}. 

\begin{figure}[tp]   
\includegraphics[width=\columnwidth]{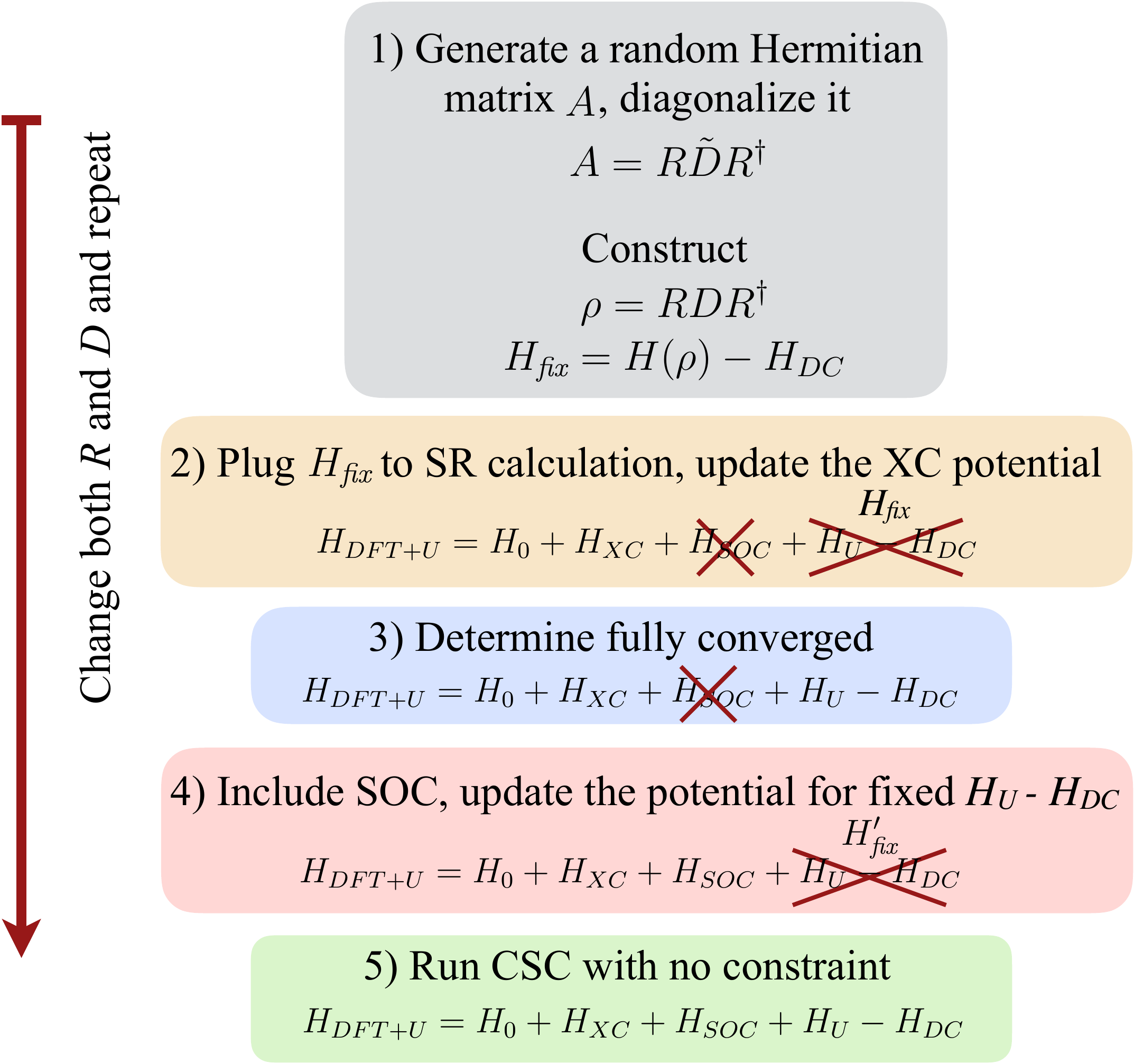}
\caption{A schematic representation of the RDMC scheme used in this work.} 
\label{fig:recipe}
\end{figure}

\subsection{Random density matrix control}
\label{instruction}
A schematic representations of our generalized OMC scheme, Random Density Matrix Control (RDMC), is shown in Fig.~\ref{fig:recipe}. The main idea is to use constrained random density matrices to construct the initial local Hartree-Fock potential, and in this way get a representative sampling of the total energy landscape. In addition, to have a better control over the convergence and trace how each simulation gets attracted to the emerging local minima, the simulations are converged in steps. The steps alternate between adding a new term to the Hamiltonian (\ref{eqn:h}) and update the local Hartree-Fock potential $H_U - H_{DC}$. 

Let us go through the RDMC scheme in Fig.~\ref{fig:recipe} in more detail:
\begin{enumerate}
\item Setup: Generate a random unitary transformation $R$ from the eigenvectors of a random (complex) hermitian matrix $A$. In the case of a scalar-relativistic (SR) starting point, $A$ is restricted to be spin-block diagonal. $R$ represents the unitary transformation that diagonalizes the initial density matrix.

Next, construct the diagonal occupation matrix $D$ with a total occupation set to the nominal occupation, {\em i.e.} 6 for FeO and 2 for UO$_2$. To cover all the corner cases, set the orbital occupations to either 0 or 1, as in the OMC scheme. The interior points can be sampled uniformly through the use of the randfixsum algorithm~\cite{randfixedsum}\footnote{In the present work we sampled the interior in a slightly non-uniform way. If a spin channel is less than half-filled, $N_{\sigma} < M/2$ where $M$ is the number of orbitals, we simply set the occupation of the corresponding spin-orbitals to a random number uniformly distributed between 0 and 1, and after that rescale all the elements to obtain the given filling $N_{\sigma}$. The scaling bias the sampling away from the already covered corner cases. If the spin channel is set to be more than half-filled ($N_{\sigma} > M/2$) we restrict the initial distribution to be between $N_{\sigma}/M$ and 1, where M is the number of orbitals, before the rescaling of the elements. This restriction gradually excludes points close to the corners, {\em i.e.} simulations with almost empty orbitals, as the filling of the spin-channel becomes larger. All matrices $D$ with elements that become larger than 1 after the rescaling were discarded.}. Additional control of the seeding procedure can be gained by scanning over selected regions of the spin moment.

Combine the diagonal matrix $D$ and the unitary matrix $R$ to form a random but constrained density matrix $\rho_i = R^\dagger D R$. The conventional OMC starting points are added to the sample by directly setting $\rho_i = D$ for the integer valued D matrices. 

Finally, generate the corresponding local Hartree-Fock potential $H_\textit{fix} = H_U[\rho_i]-H_{DC}[\rho_i]$. 

\item Converge the SR simulations using the fixed local Hartree-Fock potential $H_\textit{fix}$. This is to suppress the large initial fluctuations in the orbital occupations that otherwise may shift the solution far from the initial starting point. 

\item Allow the local potential converge to a self-consistent solution using the SR scheme.

\item Include SOC in the DFT+$U$ calculation, but keep the local potential $H'_\textit{fix}$ of the previous step fixed. 

\item Allow the DFT+$U$ calculation to converge without any constraints.
\end{enumerate}

Step 2 and 3 are only used when SOC acts as a weak perturbation, {\em e.g.} in the TMOs. Including SOC in the Hamiltonian already from the start did not reveal any additional minima. For UO$_2$ we go directly to step 4 after the initial setup. The scan over the spin moment in step 1, both for the random density matrix and the OMC points, was set to cover the interval 0.2 to its maximal value for the TMOs, while for UO$_2$ we selected the starting configurations with $n_{\uparrow} = 2$ and $n_{\downarrow} = 0$.

The successful benchmark of the RDMC scheme applied to UO$_2$ is presented in Appendix~\ref{app:benchmark}.

\section{Results and discussions}
\label{result}
\subsection{NiO}
\begin{figure}[tp]   
\includegraphics[width=\columnwidth]{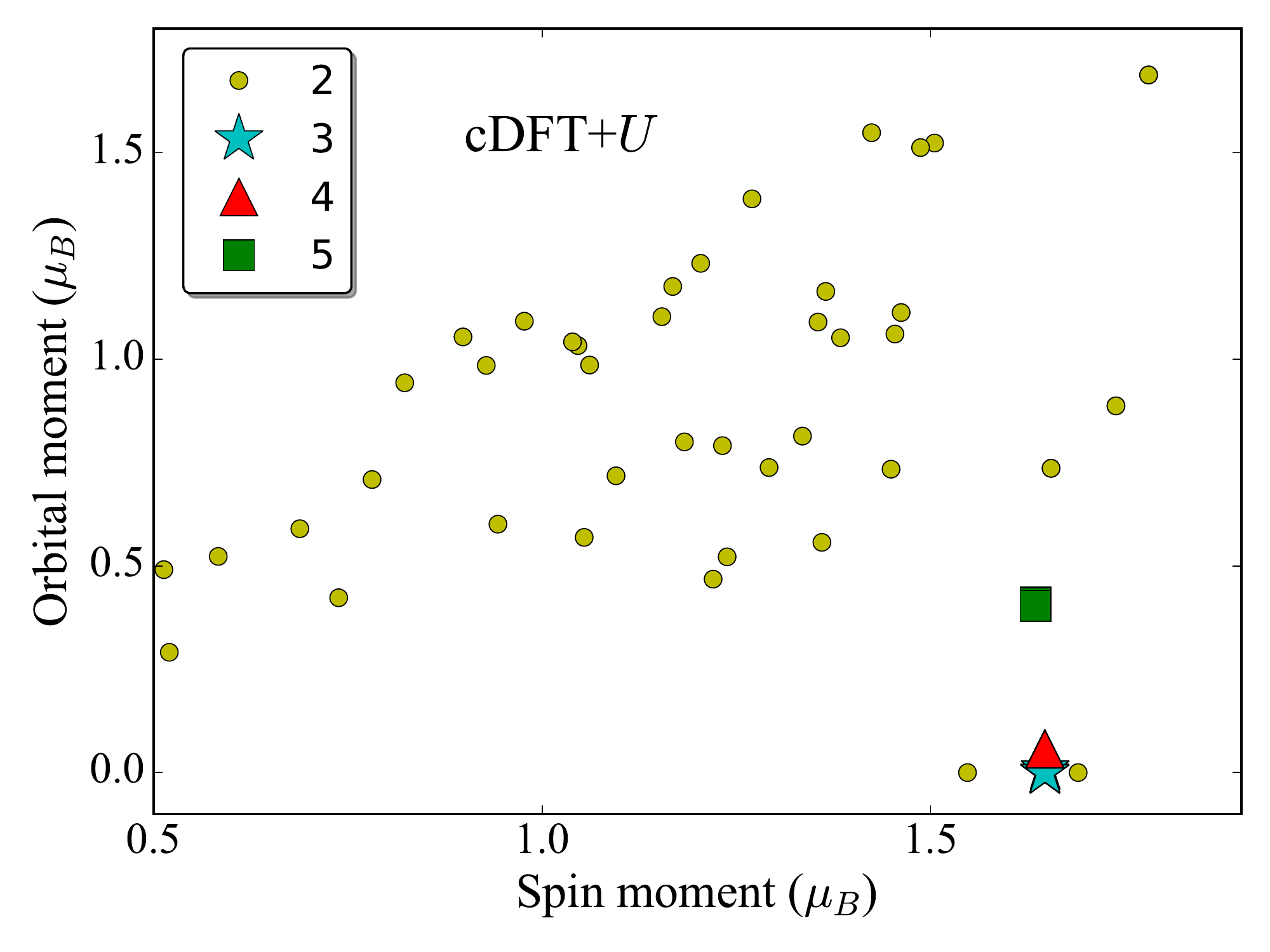}
\includegraphics[width=\columnwidth]{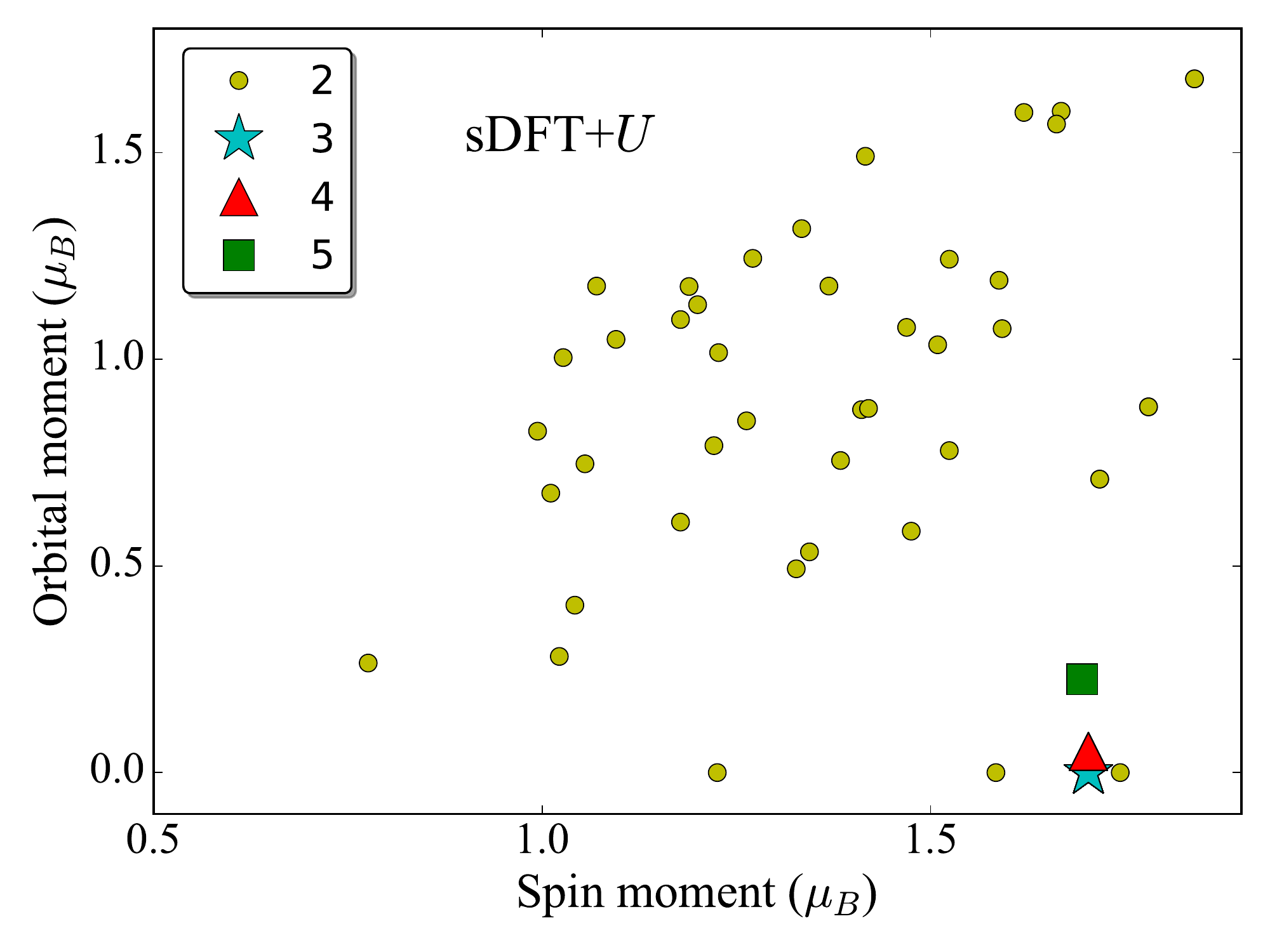}
\caption{Orbital vs. spin moment of NiO after steps 2 to 5 in the RDMC scheme. The upper and lower panel show cDFT+$U$ and sDFT+$U$, respectively.}
\label{fig:nio-lda-lsda}
\end{figure}

Anti-ferromagnetic NiO has a rhombohedral structure with the space group $R\bar{3}m$~\cite{feo1}. The lattice constant is about 4.18\AA~and the magnetisation axis is along the [111] direction. The Ni 3d orbitals split into $e_{g}$ and $t_{2g}$-like states due to the strong hybridization with the O 2p orbitals. Considering the occupations of eight electrons in the $d$ shell, in the high-spin state the $t_{2g}$-like orbitals become filled and the $e_g$ orbitals become maximally spin-polarized. A band gap forms in NiO already at the DFT level, although its size is underestimated~\cite{kamal}. The $U$ correction improves the size of the gap. 

Using the RDMC scheme, we calculate the local Ni $3d$ orbital and spin moment with both the cDFT+$U$ and the sDFT+$U$ approach, as shown in Fig.~\ref{fig:nio-lda-lsda}. The yellow circles are the moments obtained after step 2 of the RDMC scheme, {i.e.} with a fixed random seeding potential that produces a wide range of spin and orbital moments. Next, the stars represent the moments obtained by allowing the $H_U$ potential to relax (step 3). Note that in this step the spin and orbital moment of all the points converge to only two separate values, representing a high-spin and a non-magnetic solution (not shown). The high-spin moments are quite similar in cDFT+$U$ (1.63$\mu_B$) and sDFT+$U$ (1.67$\mu_B$). The red triangles correspond to the moments after SOC has been added while the $H_U - H_{DC}$ potential is kept fix (step 4). The addition of SOC makes a non-zero orbital moment favorable, but the fixed local Hartree-Fock potential keeps the orbital moments small. Finally, the green squares represent the fully self-consistent moments including SOC (step 5). The update of the $H_U - H_{DC}$ potential in the presence of SOC gives sizable orbital moments of about {$0.41 \mu_B$} in cDFT+$U$ and {$0.23 \mu_B$} in sDFT+$U$. All the high-spin simulations end up in a single sharp minimum, with a total energy spread of about 1 meV. The low-spin solutions get instead trapped in a broad meta-stable valley about 1 eV higher in energy than the high-spin solutions.  

\begin{figure}[tp]   
\includegraphics[width=\columnwidth]{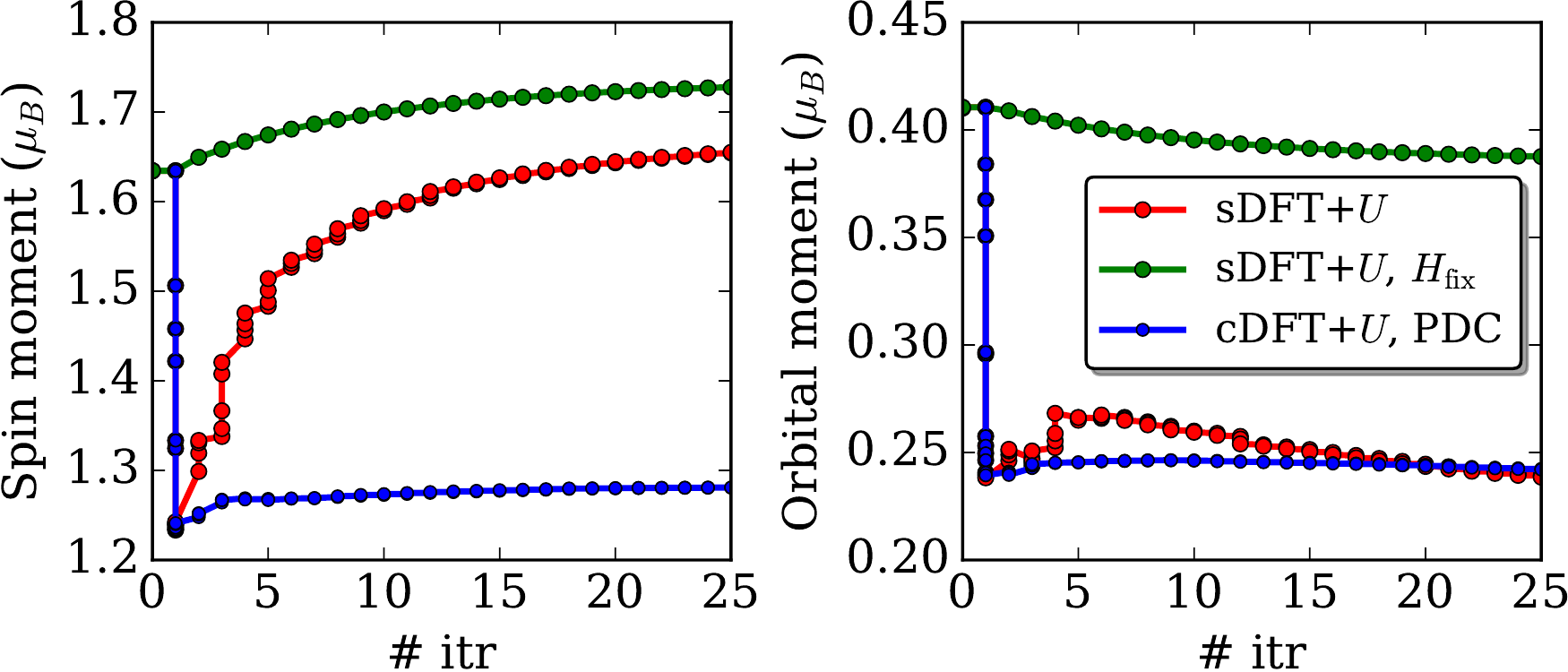}
\caption{Spin and orbital moments of NiO vs the iterations number, starting from a converged cDFT+$U$ simulation. Three different methods are used: sDFT+$U$ (red line), sDFT+$U$ but with fixed $H_U - H_{DC}$ (green line), and cDFT+$U$ using spin-polarized DC (blue line).}
\label{fig:nio-monit}
\end{figure}
The combined spin and orbital moments give a Ni magnetic moment of {$2.04 \mu_B$} and {$1.90 \mu_B$} in cDFT+$U$ and sDFT+$U$, respectively. Both of these values are within the broad range {1.8 -- 2.2 $\mu_B$} of reported experimental magnetic moment\cite{alperin1961,fernandez1998,rinaldi2016}, with the value of cDFT+$U$ particularly close to the reported value $2.02 \pm 0.04 \mu_B$ of a recent Neutron diffraction study\cite{brok2015}.

The difference in the orbital moment of the high-spin cDFT+$U$ and sDFT+$U$ solutions can be due to either the XC functional or the double counting. 
To gain additional information about the underlying mechanism, {\em i.e.} what causes the smaller orbital moment in sDFT+$U$ as compared to cDFT+$U$, we consider three different simulations starting from the same converged cDFT+$U$ solution. The first simulation is ordinary sDFT+$U$, the second sDFT+$U$ but with $H_U - H_{DC}$ kept fixed to its starting cDFT+$U$ values (sDFT+$H_{fix}$), and the third cDFT+$U$ but with the spin-polarized $H_{DC}$ of sDFT+$U$ (cDFT+$U$+PDC). To further unravel the effect of the double counting and the XC potential we convergence $H_U$ before updating $H_{XC}$, which corresponds to the vertical lines in the results shown in Fig.~\ref{fig:nio-monit}. As can be seen, the spin-polarized DC suppresses the spin and orbital moments significantly already in the very first iteration (dark blue and light red lines). The reduction of the orbital moment follow from the effective reduction of SOC at lower spin moments, which plays an important role in NiO as seen in the difference between step 3 and step 5 in Fig.~\ref{fig:nio-lda-lsda}. 
The LSDA potential with $H_{fix}$, on the other hand, change the moments only by about a few percent (green lines) by increasing the spin moment and decreasing the orbital moment. The sDFT+$U$ solution manage to recover the spin moment at self-consistency, but its orbital moment stays suppressed close to the value of cDFT+$U$+PDC. Therefore, we conclude that in NiO the suppression of the orbital moment in sDFT+$U$ is effectively due to the interplay between SOC and the DC potential rather than the spin-polarized XC functional.

\subsection{CoO}
\begin{figure}[tp]   
\includegraphics[width=\columnwidth]{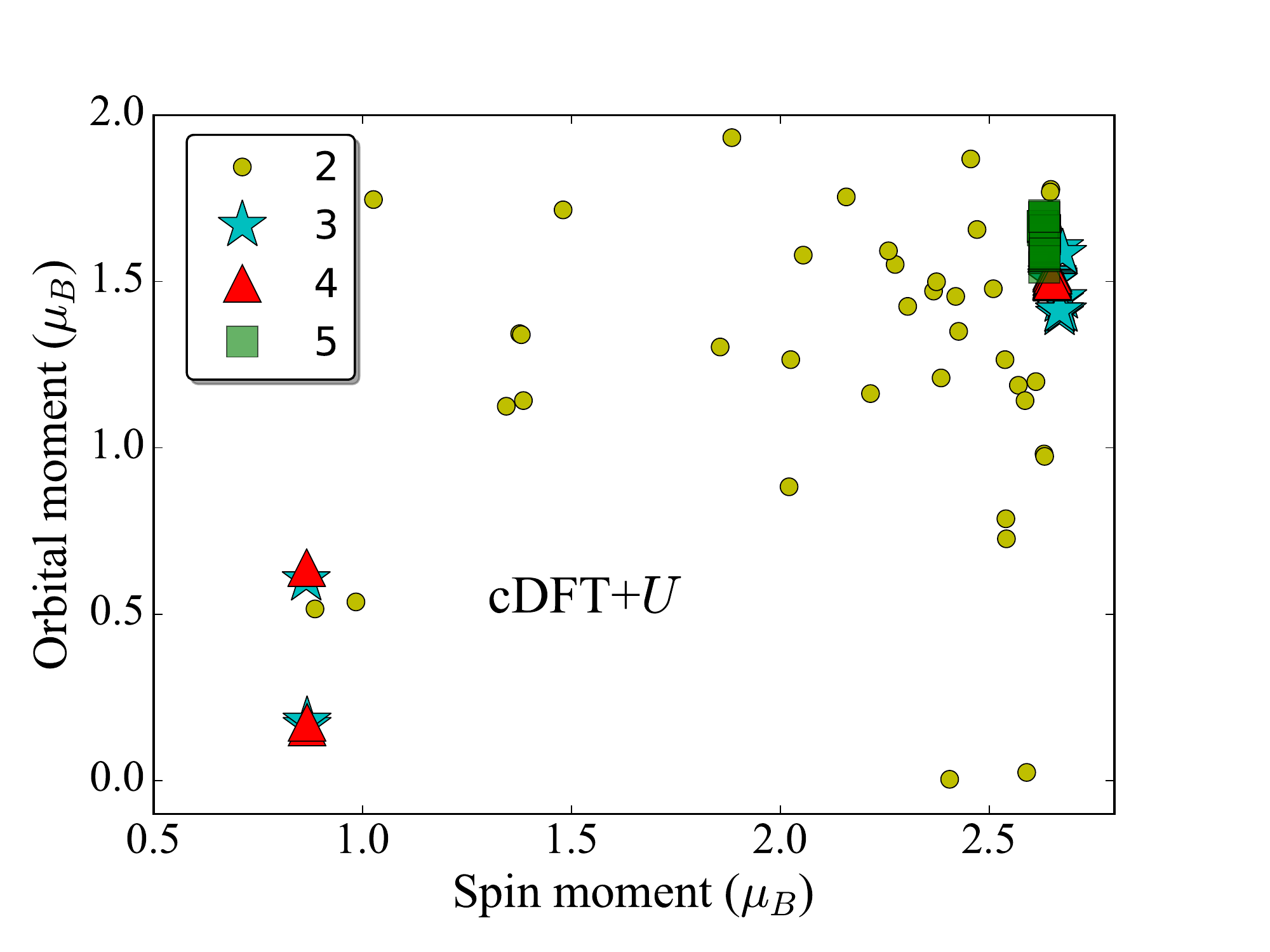}
\includegraphics[width=\columnwidth]{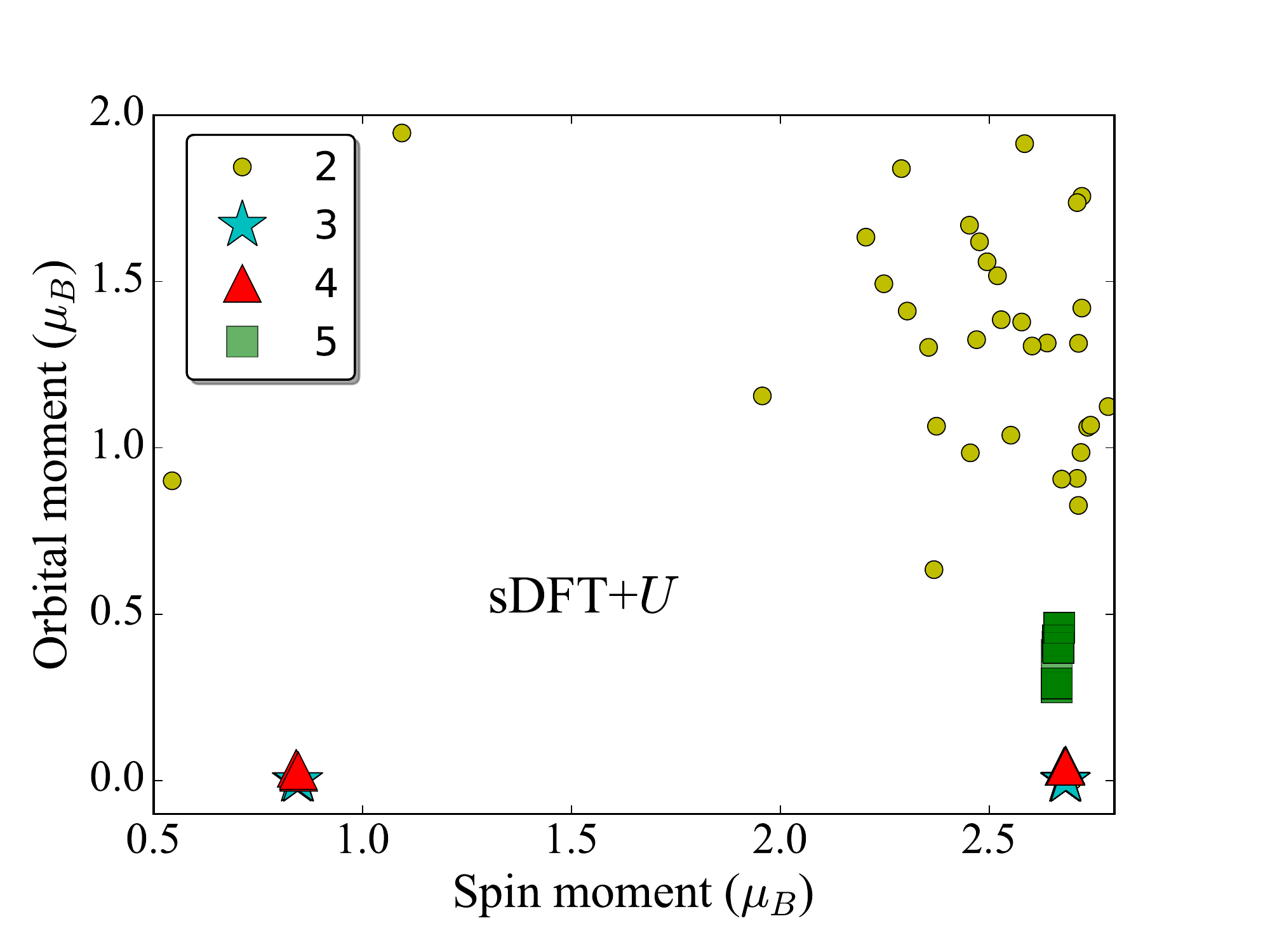}
\caption{Orbital vs. spin moment of CoO after steps 2 to 5 in the RDMC scheme. The upper and lower panel show cDFT+$U$ and sDFT+$U$, respectively.} 
\label{fig:coo-lda-lsda}
\end{figure}

\begin{figure}[tp]   
\includegraphics[width=\columnwidth]{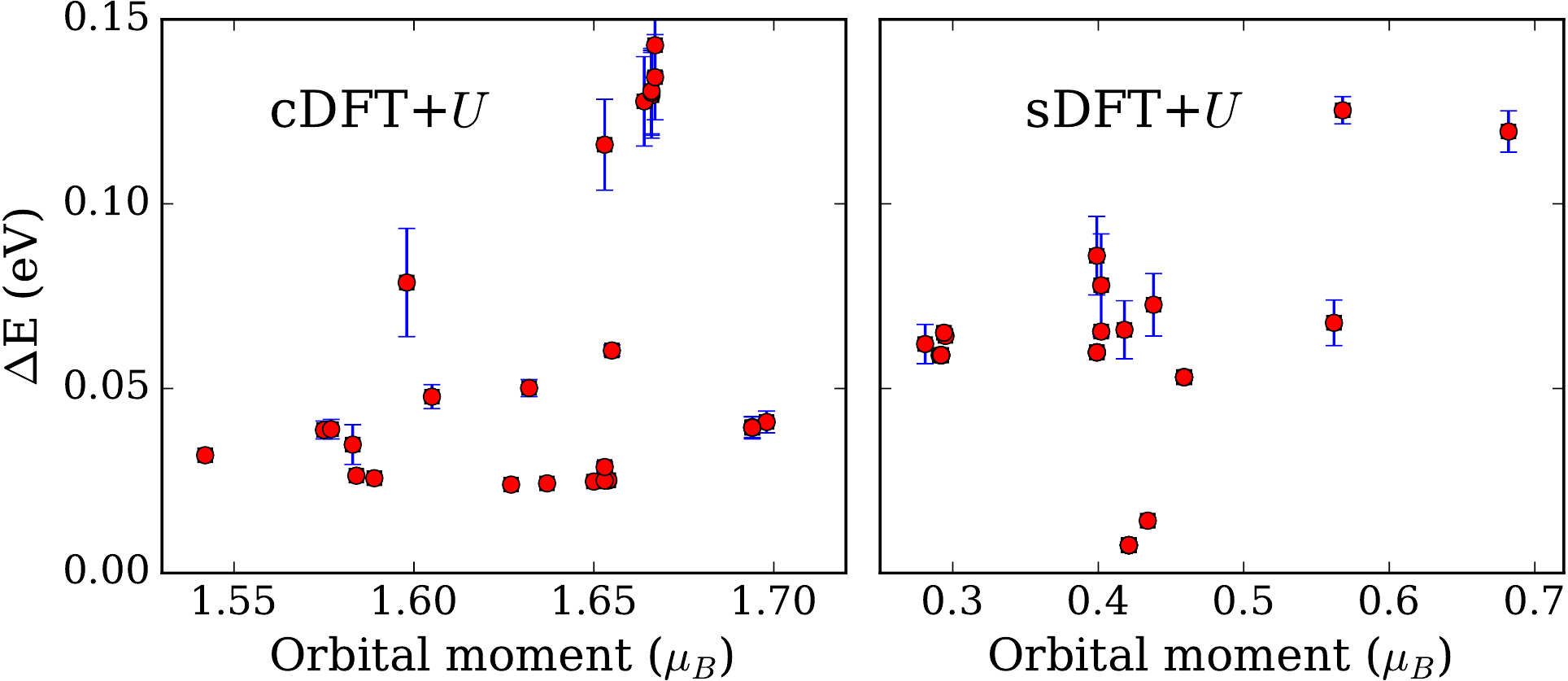}
\caption{The total energy vs Co 3d orbital moment of the fully self-consistent high-spin solutions. The error bars represent the spread in energy over the previous 10 iterations.}
\label{fig:coo-energy}
\end{figure}

The orbital occupation in CoO favors an orthorhombic distortion that reduces the symmetry of the system to monoclinic~\cite{kamal,schron,solov,feo2}. The primitive unitcell consists of four formula units, {\em i.e.} Co$_4$O$_4$, and the experimental lattice constant is 4.26{\AA}. Fig.~\ref{fig:coo-lda-lsda} shows the spin moment vs orbital moment of the solutions found using the RDMC method. In the absence of SOC (step 3) several different local minima emerge. In addition to the high-spin solution also local minima with intermediate spin is found both in cDFT+$U$ and sDFT+$U$. As in NiO, both methods give similar spin moments, but sDFT+$U$ clearly suppresses the orbital moment compared to cDFT+$U$. The intermediate spin states become unstable when SOC is added (step 4 and 5) and all the corresponding simulations converge instead to the high-spin solution. The role of SOC for the Co orbital moment in cDFT+$U$ is much smaller compared to the role it played in NiO, as seen in the almost overlapping distributions of orbital moment in step 3 and step 5 in the upper panel of Fig.~\ref{fig:coo-lda-lsda}. The total energy data in Fig.~\ref{fig:coo-energy} shows that the converged high-spin solutions in cDFT+$U$ form parabolic shapes with respect to the orbital moment, while in sDFT+$U$ the points follow no clear pattern. Nevertheless, the spread in total energy is around 150 meV in both cases.

\begin{figure}[tp]   
\includegraphics[width=\columnwidth]{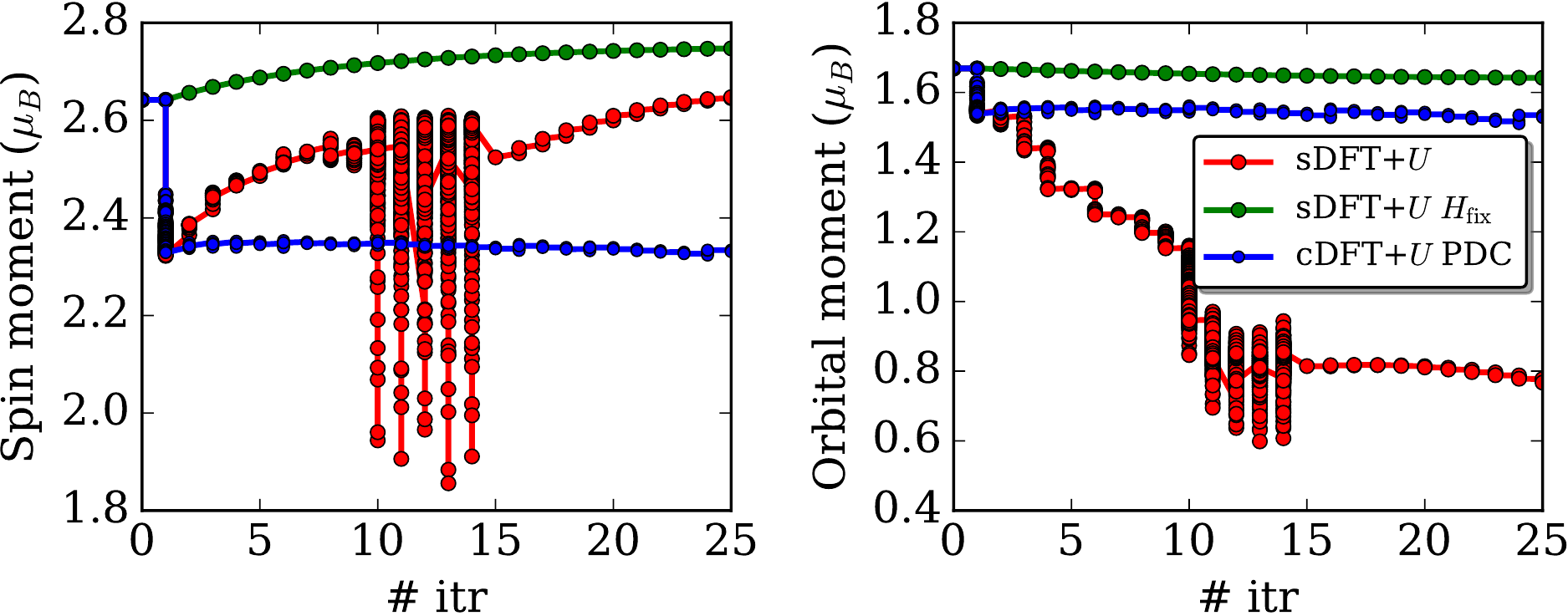}
\caption{Spin and orbital moments of CoO vs the iterations number, starting from a converged cDFT+$U$ simulation. Three different methods are used: sDFT+$U$ (red line), sDFT+$U$ but with fixed $H_U - H_{DC}$ (green line), and cDFT+$U$ using spin-polarized DC (blue line).}
\label{fig:coo-monit}
\end{figure}
A comparison of sDFT+$U$, sDFT+$H_{fix}$, and cDFT+$U$+PDC starting from a high-spin cDFT+$U$ solution is shown in Fig.~\ref{fig:coo-monit}. The spin-polarized DC initially suppress both the spin and orbital moment from their initial values, but in contrast to NiO by only a few percent. The small drop in orbital moment is consistent with an effective mixing of the unstable intermediate spin and the high-spin solutions shown in Fig.~\ref{fig:coo-energy}, due to the artificial suppression of the spin moment in the first iteration. As the sDFT+$U$ converge to its high-spin solution its spin moment recovers, but the orbital moment continues instead to drop until it eventually reaches about 0.5 $\mu_B$, a 70\% decrease compared to the cDFT+$U$ starting point. The orbital moment in sDFT+$H_{fix}$ is only weakly suppressed by the LSDA potential, and the already almost maximally polarized spin moment is only slightly enhanced. Therefore, we conclude that in CoO, in contrast to the NiO, that the suppression of the orbital moment in sDFT+$U$ is effectively due to the spin-polarized XC functional rather than the DC potential.

Finally, we mention that the obtained spin moment of 2.63 (2.67) $\mu_B$ in cDFT+$U$ (sDFT+$U$) is in a good agreement by the theoretical value reported in Ref.~\cite{solov,schron}. The orbital moment value, depending on the XC functional, varies between 1.5-1.8$\mu_B$ (0.3-0.5$\mu_B$) in the cDFT+$U$ (sDFT+$U$) approach. The total moment in cDFT+$U$ compares rather well with the experimental magnetic moments of 3.8-4.0$\mu_B$~\cite{feo1,coo-exp1}.

\subsection{FeO}
\begin{figure}[tp]   
\includegraphics[width=\columnwidth]{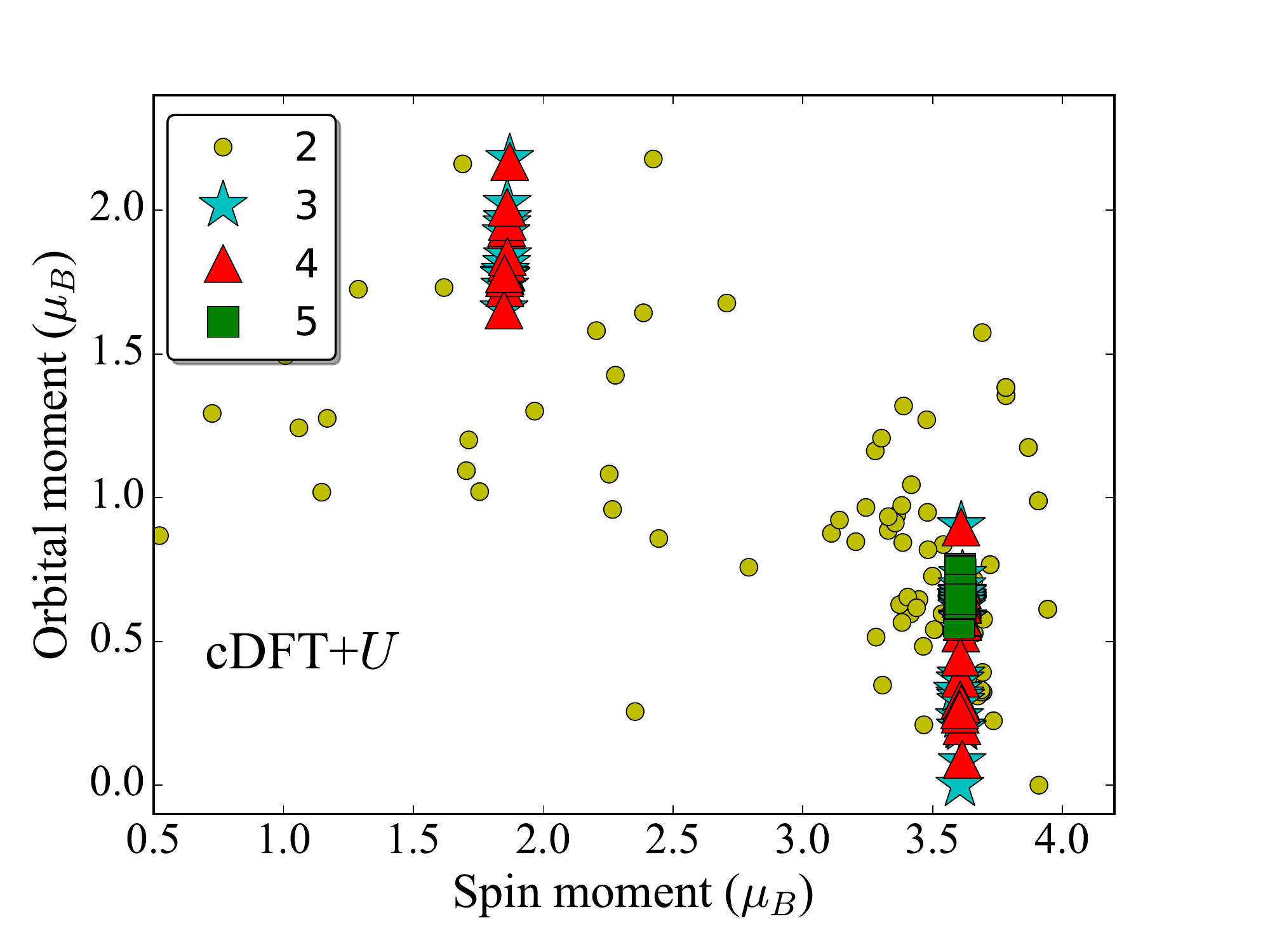}
\includegraphics[width=\columnwidth]{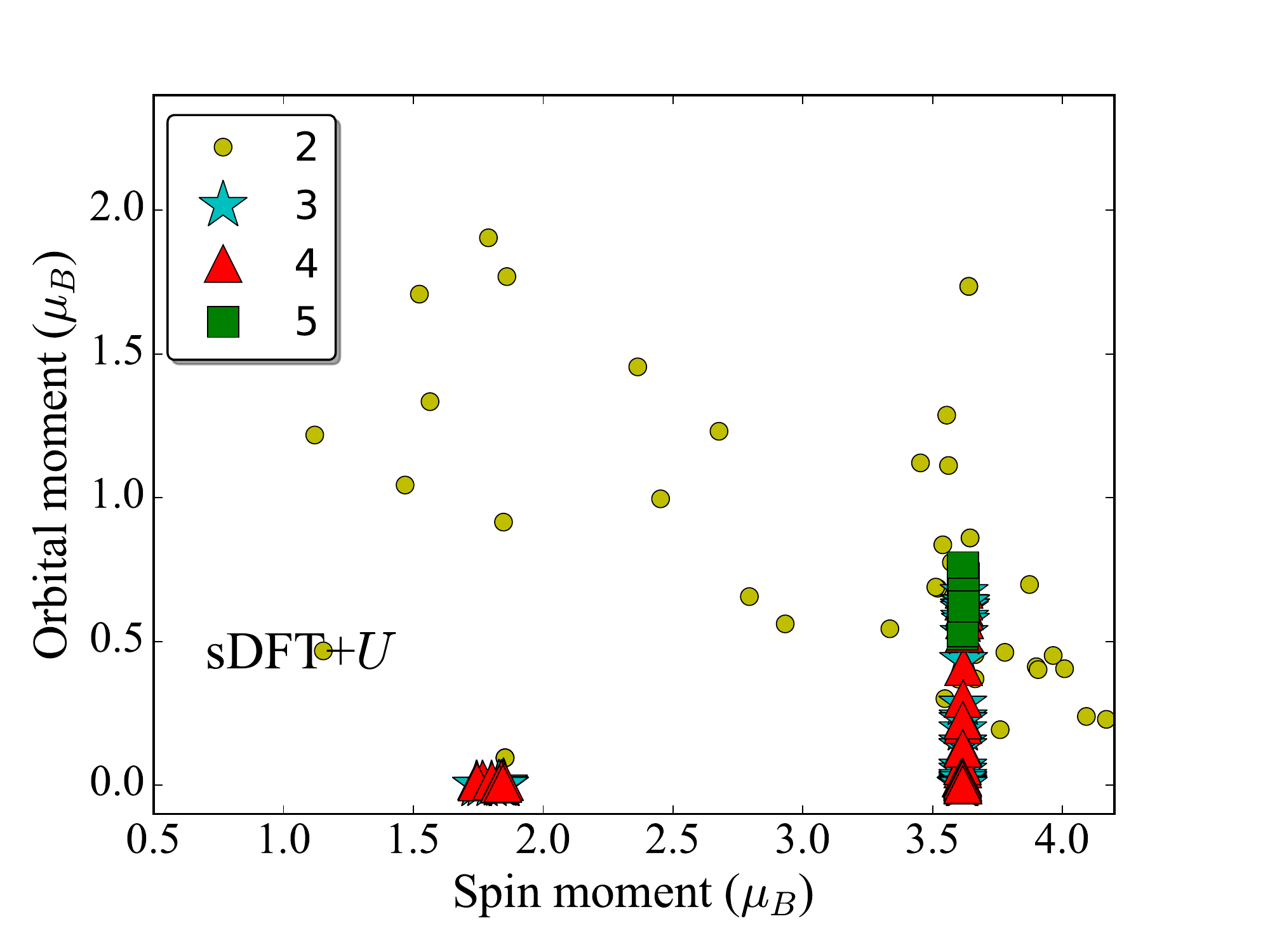}
\caption{Orbital vs. spin moment of FeO after steps 2 to 5 in the RDMC scheme. The upper and lower panel show cDFT+$U$ and sDFT+$U$, respectively.} 
\label{fig:feo-lda-lsda}
\end{figure}

Similar to CoO, the orbital ordering in FeO introduces an orthorhombic distortion which reduce the symmetry of the system to monoclinic~\cite{kamal,schron,solov,feo2}. This yields a primitive unit cell containing four formula units, {\em i.e.} Fe$_4$O$_4$, with the experimental lattice constant 4.33{\AA}. FeO is an anti-ferromagnetic metal in DFT with the LSDA functional, but the addition of the local Hartree-Fock potential in DFT+$U$ yields an insulator. However, several different calculated spin and orbital moments, as well as band gap characters, have been reported in the literature~\cite{mazin,kamal,schron,solov} which is a strong indication of several local minima in the total energy landscape. 

\begin{figure}[tp]   
\includegraphics[width=\columnwidth]{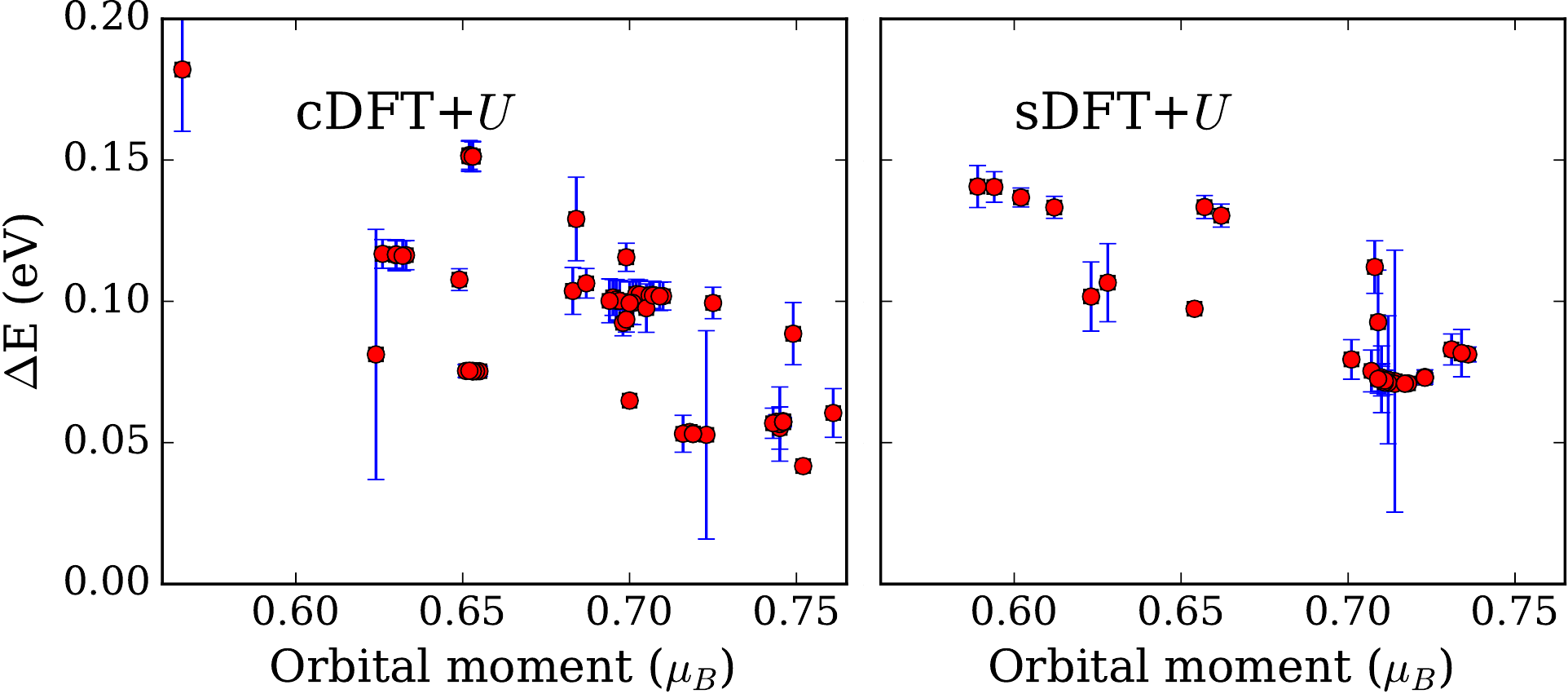}
\caption{The total energy vs Fe 3d orbital moment for the high-spin solutions. The error bars represent the spread in energy over the previous 10 iterations.}
\label{fig:feo-energy}
\end{figure}
\begin{figure}[tp]   
\includegraphics[width=\columnwidth]{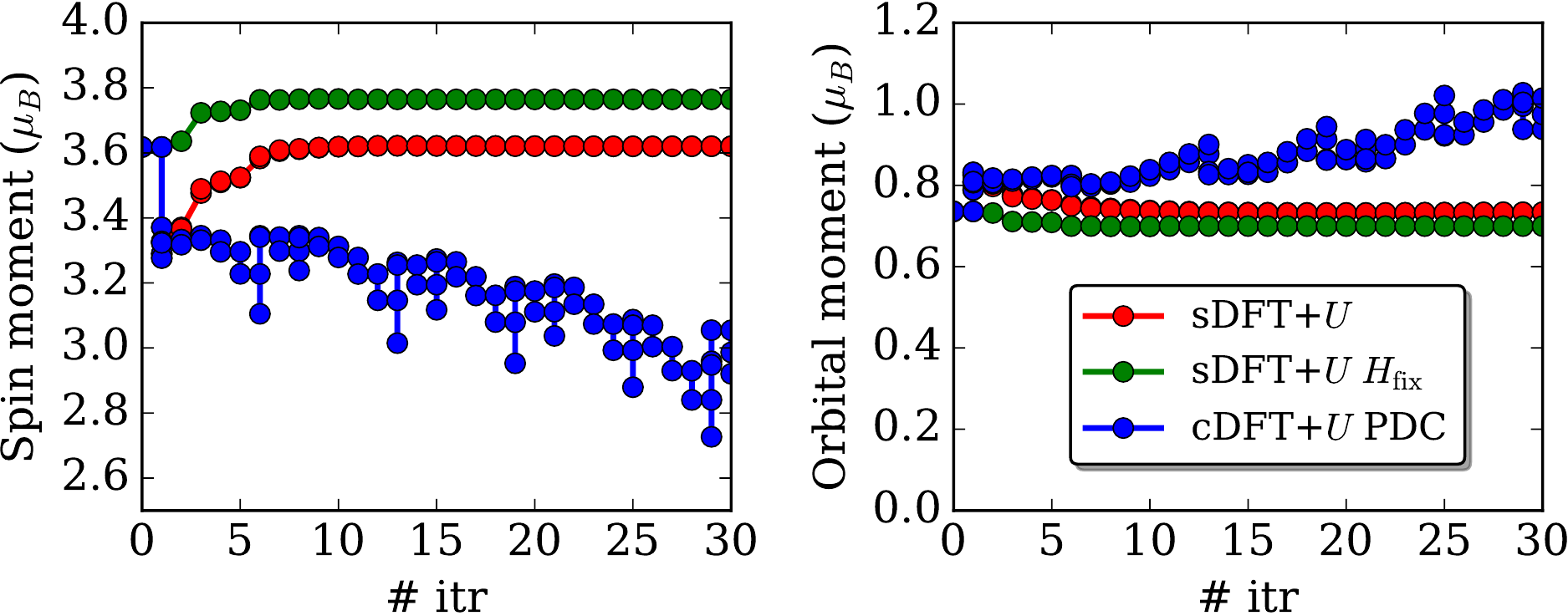}
\caption{Spin and orbital moments of FeO vs the iterations number, starting from a converged cDFT+$U$ simulation. Three different methods are used: sDFT+$U$ (red line), sDFT+$U$ but with fixed $H_U - H_{DC}$ (green line), and cDFT+$U$ using spin-polarized DC (blue line).}
\label{fig:feo-monit}
\end{figure}
Indeed, the RDMC scheme applied to FeO yields both high-spin and intermediate-spin solutions in the absence of SOC (step 3), as seen in Fig.~\ref{fig:feo-lda-lsda}. Interestingly, the orbital moment of the intermediate-spin states is strongly enhanced in cDFT+$U$, but completely suppressed in sDFT+$U$. With the addition of SOC (step 5) the intermediate-spin states become unstable and converge instead to the high-spin minimum. However, as shown in Appendix~\ref{app:kpoints}, the smooth convergence depends on the number of the $k$-points. A less dense $k$-mesh of 24$\times$24$\times$24 results in the trapping of some solutions in the intermediate spin region. In FeO the orbital moment of the high-spin solution in cDFT+$U$ and sDFT+$U$ are quite similar, in contrast to the substantial differences seen for NiO and CoO. The addition of SOC does not strictly shift the orbital moment of the high-spin solutions but rather makes the low moment minima unstable. These results show that the reduction of the orbital moment in sDFT+$U$ is not universal, but depends on the details of the electronic structure.

The high-spin solutions span an energy range of about 150 eV in cDFT+$U$ and 100 meV in sDFT+$U$, as shown plotted against the orbital moment in Fig.~\ref{fig:feo-energy}. In contrast to CoO, here the cDFT+$U$ data is rather erratic, with a large number of clustered minima, while most of the sDFT+$U$ simulations cluster around a parabolic minimum centred at 0.72 $\mu_B$. This high-lights the strong material dependence of the local minima, and the need for a general method, such as RDMC, to fully explore the formation of meta-stable states.

In FeO the orbital moment of the high-spin solution in cDFT+$U$ and sDFT+$U$ are quite similar, in contrast to the substantial differences seen for NiO and CoO. In Fig.~\ref{fig:feo-monit} we again compare sDFT+$U$, sDFT+$H_{fix}$, and cDFT+$U$+PDC, starting from a high spin cDFT+$U$ solution. The spin moment get initially reduced, as expected, by the onset of the polarized double counting, but the orbital moment is initially {\em increased}. This is again consistent with a mixing of the high-spin and intermediate spin cDFT+$U$ solutions due to the suppression of the spin moment. In sDFT+$U$ the orbital moment slowly returns close to its initial value, while the cDFT+$U$+PDC simulation is pushed towards the intermediate spin solution due to its spin suppressing double counting term. 

The obtained spin moment of 3.65$\mu_B$ and the orbital moment between 0.5-0.8$\mu_B$ are found to be in agreement with the experimental total moment of 4-4.6$\mu_B$ in Refs.~\cite{feo-exp1,feo-exp2}. The theoretical spin (orbital) moment of 3.65 (0.66)$\mu_B$ reported in Ref.~\cite{schron-mom} is one of our obtained results.

\section{Conclusions}
\label{conclusions}
In this work, we have performed a systematic study of the emergence of meta-stable states in DFT+$U$ applied to the late transition metal oxides NiO, CoO, and FeO. To this end, we extended the method of occupation matrix control by replacing the eponymous occupation matrices with constrained random density matrices. In addition we introduced a step-wise convergence of the DFT+$U$ potential to better monitor the seeding of the DFT potential and the influence of spin-orbit coupling. The resulting scheme, Random Density Matrix Control (RDMC), finds an extra local minimum in the benchmark material UO$_2$ in addition to all the meta-stable states previously reported in Ref.~\cite{amadon-uo2}. The RDMC scheme applied to the transition metal oxides without the inclusion of SOC yields both high-spin and intermediate-spin solutions in CoO and FeO, while NiO only supports a high-spin solution. While the cDFT+$U$ method based on LDA gives rise to finite orbital moments for the intermediate-spin states, the sDFT+$U$ method based on LSDA quenches their orbital moments completely. With the inclusion of SOC the intermediate spin states become unstable and the simulations converge to the high-spin solution, for which both methods yield finite orbital moments. The calculated orbital moments are larger in cDFT+$U$ compared to sDFT+$U$ for both NiO and CoO, which we traced back to be due to the polarization of the double counting and XC functional, respectively. However, both methods give very similar orbital moments for FeO, which shows that there is no universal suppression of orbital moments in sDFT+$U$.

The total magnetic moment of the cDFT+$U$ ground state of the TMOs are in good agreement with experimental data, while sDFT+$U$ produces a seemingly too small moment in the case of CoO. However, as the value of the total moment depends on the strength of the interaction parameters $U$ and $J$, and the choice of correlated orbitals, there may be other parameter regimes where sDFT+$U$ becomes more accurate.

A weakness of the RDMC scheme, which it also shares with OMC, is that the initial local Hartree-Fock potential may not be sufficient to give a representative sample of all spin and orbital moments after the initial DFT+$U$ iterations (step 2 in the RDMC scheme). This shows up in the clustering of the initial points in Figs.~\ref{fig:nio-lda-lsda}, \ref{fig:coo-lda-lsda}, and \ref{fig:feo-lda-lsda}. An interesting future research direction would be to use of additional constraining fields during these initial iterations to potentially cure this issue.

\section*{Acknowledgment}
We would like to acknowledge Igor Di Marco and Lars Nordstr{\"o}m for useful discussions and initial guidance. We acknowledge support from the Swedish Research Council (VR), eSSENCE and the KAW foundation. The computer simulations are performed on computational resources provided by NSC and UPPMAX allocated by the Swedish National Infrastructure for Computing (SNIC).

\appendix
\counterwithin{figure}{section}
\counterwithin{table}{section}

\section{UO$_2$}
\label{app:benchmark}
\begin{figure}[tp]   
\includegraphics[width=\columnwidth]{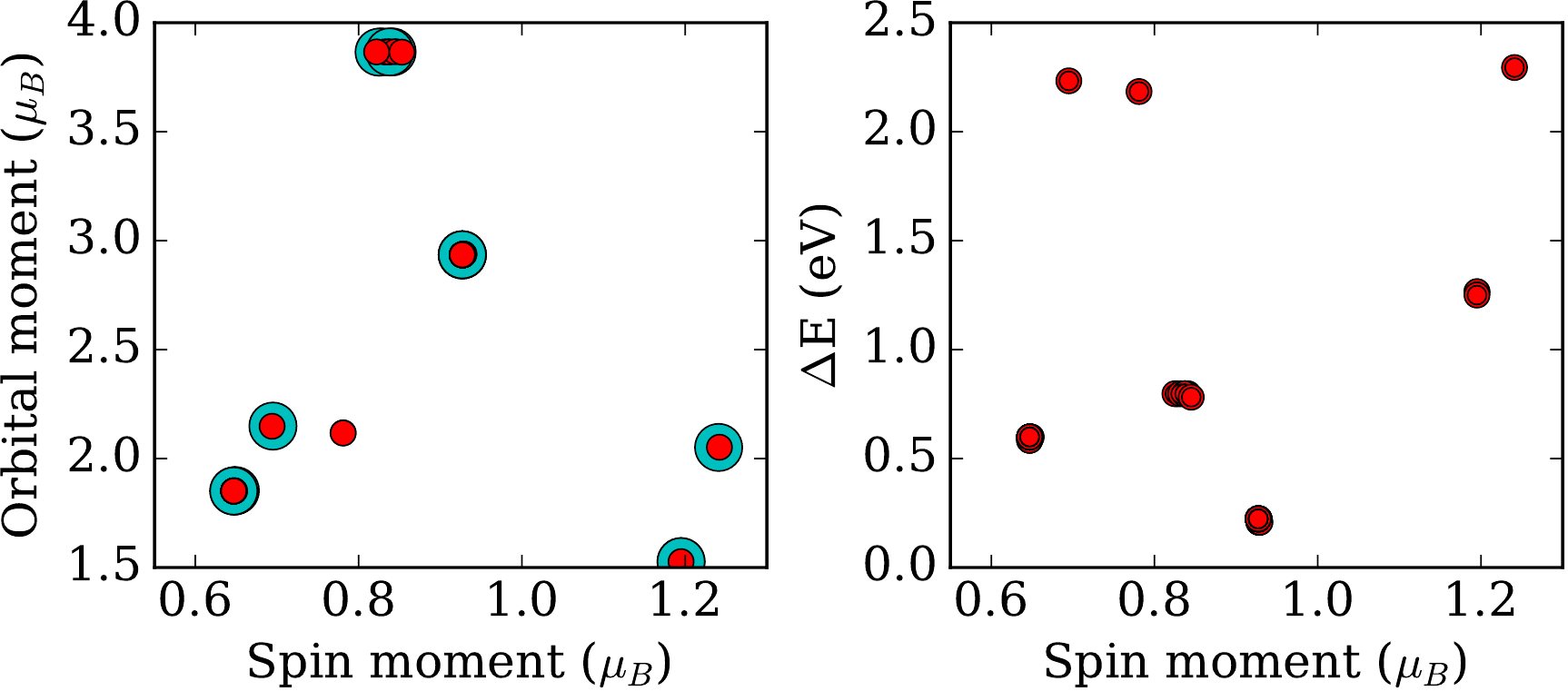}
\caption{Orbital moment (upper panel) and the energy difference (lower panel) vs the spin moment of UO$_2$ using sDFT+$U$. The crosses mark the subset of the RDMC solutions found by using diagonal density matrices with integer occupation as in OMC.}
\label{fig:uo2benchmark}
\end{figure}
The RDMC scheme includes by construction all the seeding points produced by the OMC scheme. It should hence find at all the minima located by OMC, but can potentially unveil additional solutions. Anti-ferromagnetic (1{$\mathbf{k}$}) UO$_2$ serves as an excellent benchmark to compare the two methods as it presents a large number of local minima~\cite{amadon-uo2,u-ramping}. The system has a cubic crystal symmetry (O$_h$), but the 1{$\mathbf{k}$} AMF order along the (001) direction reduces the overall symmetry to D$_{4h}$. A direct comparison between the two schemes using cDFT+$U$ is presented in Fig.~\ref{fig:uo2benchmark}. About 60 random density matrix seeds were used in addition the to OMC points. The OMC results, marked with large blue circles, are in good agreement with the ones reported in Ref.~\cite{amadon-uo2}. 
The main difference between the RDMC and OMC schemes at the eV scale is that the RDMC scheme is able to capture an additional local minimum with the spin and orbital moment of 0.78$\mu_B$ and 2.2$\mu_B$, respectively. This additional solution shows that it is in general necessary to use a random seeding, even in a highly symmetric $4f$ material, to fully map out the energy landscape. 

There are also differences on the meV scale between OMC and RDMC within each local energy minima. For example, the ground state produced by the random seedings has 13 meV lower energy than the lowest energy state found by the integer occupation matrix seeding. However, the spin and the orbital moment are well-captured in both ways of seedings with the spin (orbital) moment of 0.92$\mu_B$ (2.93$\mu_B$). 

\section{$k$-point convergence}
\label{app:kpoints}
\begin{figure}[bp]   
\includegraphics[width=\columnwidth]{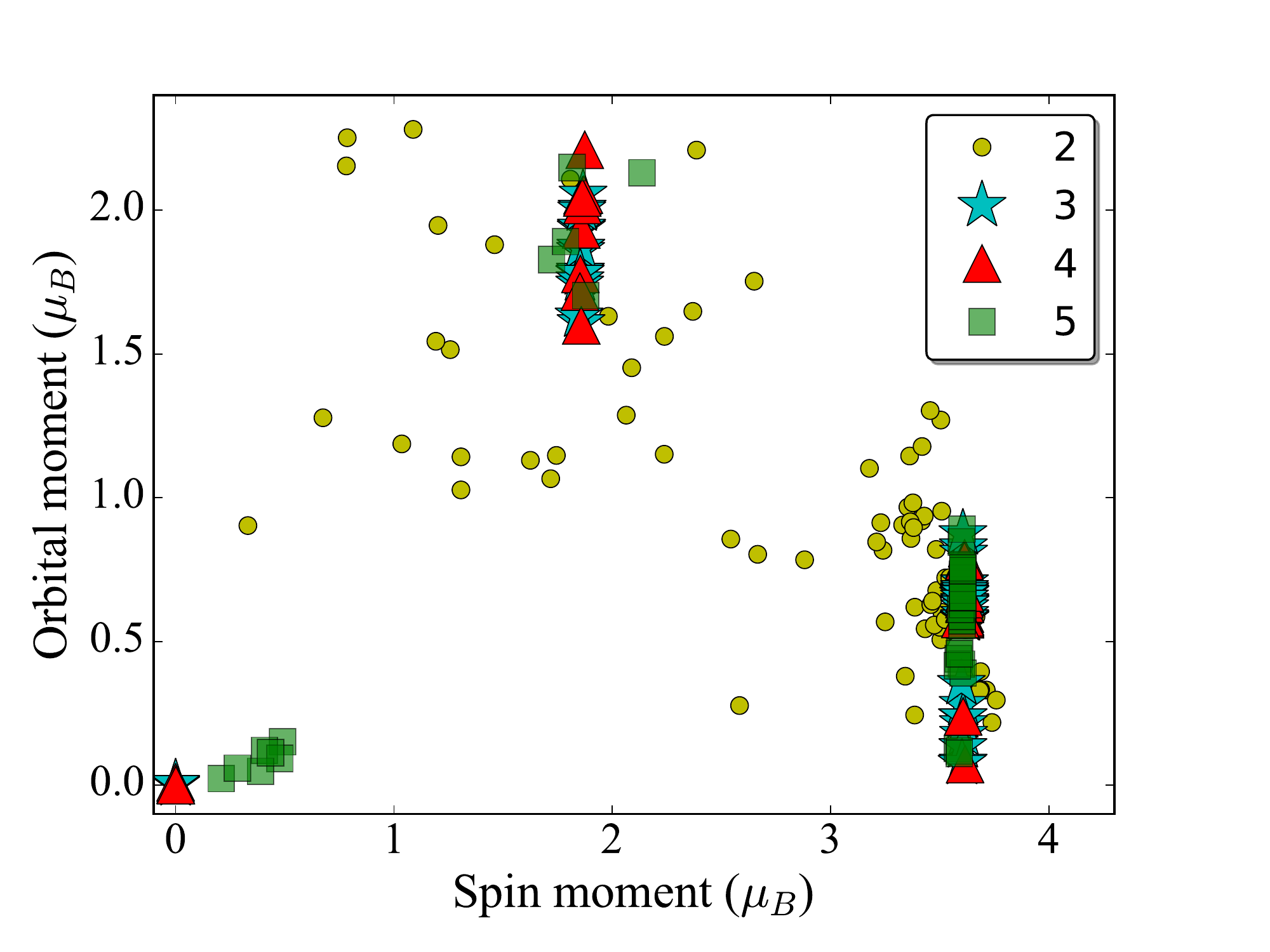}
\caption{Orbital vs. spin moment of FeO using a moderate sized $k$-mesh ($24\times24\times24$). Some of the intermediate-spin solutions are not able to converge to the high-spin minima when SOC is applied.} \label{fig:feo-lowk_vs_highk}
\end{figure}

The RDMC results for FeO with sDFT+$U$ and a $k$-point mesh of $24\times24\times24$ are shown in Fig.~\ref{fig:feo-lowk_vs_highk}.
The moderately sized $k$-point mesh and the weak SOC result in a spurious trapping of some intermediate-spin solutions, as the low temperature Fermi smearing can not correctly resolve the details of the almost flat total energy landscape.


\begin{thebibliography}{45}%
\makeatletter
\providecommand \@ifxundefined [1]{%
 \@ifx{#1\undefined}
}%
\providecommand \@ifnum [1]{%
 \ifnum #1\expandafter \@firstoftwo
 \else \expandafter \@secondoftwo
 \fi
}%
\providecommand \@ifx [1]{%
 \ifx #1\expandafter \@firstoftwo
 \else \expandafter \@secondoftwo
 \fi
}%
\providecommand \natexlab [1]{#1}%
\providecommand \enquote  [1]{``#1''}%
\providecommand \bibnamefont  [1]{#1}%
\providecommand \bibfnamefont [1]{#1}%
\providecommand \citenamefont [1]{#1}%
\providecommand \href@noop [0]{\@secondoftwo}%
\providecommand \href [0]{\begingroup \@sanitize@url \@href}%
\providecommand \@href[1]{\@@startlink{#1}\@@href}%
\providecommand \@@href[1]{\endgroup#1\@@endlink}%
\providecommand \@sanitize@url [0]{\catcode `\\12\catcode `\$12\catcode
  `\&12\catcode `\#12\catcode `\^12\catcode `\_12\catcode `\%12\relax}%
\providecommand \@@startlink[1]{}%
\providecommand \@@endlink[0]{}%
\providecommand \url  [0]{\begingroup\@sanitize@url \@url }%
\providecommand \@url [1]{\endgroup\@href {#1}{\urlprefix }}%
\providecommand \urlprefix  [0]{URL }%
\providecommand \Eprint [0]{\href }%
\providecommand \doibase [0]{http://dx.doi.org/}%
\providecommand \selectlanguage [0]{\@gobble}%
\providecommand \bibinfo  [0]{\@secondoftwo}%
\providecommand \bibfield  [0]{\@secondoftwo}%
\providecommand \translation [1]{[#1]}%
\providecommand \BibitemOpen [0]{}%
\providecommand \bibitemStop [0]{}%
\providecommand \bibitemNoStop [0]{.\EOS\space}%
\providecommand \EOS [0]{\spacefactor3000\relax}%
\providecommand \BibitemShut  [1]{\csname bibitem#1\endcsname}%
\let\auto@bib@innerbib\@empty
\bibitem [{\citenamefont {Keshavarz}\ \emph {et~al.}(2017)\citenamefont
  {Keshavarz}, \citenamefont {Kvashnin}, \citenamefont {Rodrigues},
  \citenamefont {Pereiro}, \citenamefont {Di~Marco}, \citenamefont {Autieri},
  \citenamefont {Nordstr\"om}, \citenamefont {Solovyev}, \citenamefont
  {Sanyal},\ and\ \citenamefont {Eriksson}}]{cmo}%
  \BibitemOpen
  \bibfield  {author} {\bibinfo {author} {\bibfnamefont {S.}~\bibnamefont
  {Keshavarz}}, \bibinfo {author} {\bibfnamefont {Y.~O.}\ \bibnamefont
  {Kvashnin}}, \bibinfo {author} {\bibfnamefont {D.~C.~M.}\ \bibnamefont
  {Rodrigues}}, \bibinfo {author} {\bibfnamefont {M.}~\bibnamefont {Pereiro}},
  \bibinfo {author} {\bibfnamefont {I.}~\bibnamefont {Di~Marco}}, \bibinfo
  {author} {\bibfnamefont {C.}~\bibnamefont {Autieri}}, \bibinfo {author}
  {\bibfnamefont {L.}~\bibnamefont {Nordstr\"om}}, \bibinfo {author}
  {\bibfnamefont {I.~V.}\ \bibnamefont {Solovyev}}, \bibinfo {author}
  {\bibfnamefont {B.}~\bibnamefont {Sanyal}}, \ and\ \bibinfo {author}
  {\bibfnamefont {O.}~\bibnamefont {Eriksson}},\ }\href {\doibase
  10.1103/PhysRevB.95.115120} {\bibfield  {journal} {\bibinfo  {journal} {Phys.
  Rev. B}\ }\textbf {\bibinfo {volume} {95}},\ \bibinfo {pages} {115120}
  (\bibinfo {year} {2017})}\BibitemShut {NoStop}%
\bibitem [{\citenamefont {Shick}\ \emph {et~al.}(2004)\citenamefont {Shick},
  \citenamefont {Jani\ifmmode~\check{s}\else \v{s}\fi{}}, \citenamefont
  {Drchal},\ and\ \citenamefont {Pickett}}]{pickett-uge}%
  \BibitemOpen
  \bibfield  {author} {\bibinfo {author} {\bibfnamefont {A.~B.}\ \bibnamefont
  {Shick}}, \bibinfo {author} {\bibfnamefont {V.}~\bibnamefont
  {Jani\ifmmode~\check{s}\else \v{s}\fi{}}}, \bibinfo {author} {\bibfnamefont
  {V.}~\bibnamefont {Drchal}}, \ and\ \bibinfo {author} {\bibfnamefont {W.~E.}\
  \bibnamefont {Pickett}},\ }\href {\doibase 10.1103/PhysRevB.70.134506}
  {\bibfield  {journal} {\bibinfo  {journal} {Phys. Rev. B}\ }\textbf {\bibinfo
  {volume} {70}},\ \bibinfo {pages} {134506} (\bibinfo {year}
  {2004})}\BibitemShut {NoStop}%
\bibitem [{\citenamefont {Mazin}\ and\ \citenamefont {Anisimov}(1997)}]{mazin}%
  \BibitemOpen
  \bibfield  {author} {\bibinfo {author} {\bibfnamefont {I.~I.}\ \bibnamefont
  {Mazin}}\ and\ \bibinfo {author} {\bibfnamefont {V.~I.}\ \bibnamefont
  {Anisimov}},\ }\href {\doibase 10.1103/PhysRevB.55.12822} {\bibfield
  {journal} {\bibinfo  {journal} {Phys. Rev. B}\ }\textbf {\bibinfo {volume}
  {55}},\ \bibinfo {pages} {12822} (\bibinfo {year} {1997})}\BibitemShut
  {NoStop}%
\bibitem [{\citenamefont {Amadon}\ \emph {et~al.}(2008)\citenamefont {Amadon},
  \citenamefont {Jollet},\ and\ \citenamefont {Torrent}}]{amadon-ce}%
  \BibitemOpen
  \bibfield  {author} {\bibinfo {author} {\bibfnamefont {B.}~\bibnamefont
  {Amadon}}, \bibinfo {author} {\bibfnamefont {F.}~\bibnamefont {Jollet}}, \
  and\ \bibinfo {author} {\bibfnamefont {M.}~\bibnamefont {Torrent}},\ }\href
  {\doibase 10.1103/PhysRevB.77.155104} {\bibfield  {journal} {\bibinfo
  {journal} {Phys. Rev. B}\ }\textbf {\bibinfo {volume} {77}},\ \bibinfo
  {pages} {155104} (\bibinfo {year} {2008})}\BibitemShut {NoStop}%
\bibitem [{\citenamefont {Anisimov}\ \emph {et~al.}(1991)\citenamefont
  {Anisimov}, \citenamefont {Zaanen},\ and\ \citenamefont {Andersen}}]{ldau1}%
  \BibitemOpen
  \bibfield  {author} {\bibinfo {author} {\bibfnamefont {V.~I.}\ \bibnamefont
  {Anisimov}}, \bibinfo {author} {\bibfnamefont {J.}~\bibnamefont {Zaanen}}, \
  and\ \bibinfo {author} {\bibfnamefont {O.~K.}\ \bibnamefont {Andersen}},\
  }\href {\doibase 10.1103/PhysRevB.44.943} {\bibfield  {journal} {\bibinfo
  {journal} {Phys. Rev. B}\ }\textbf {\bibinfo {volume} {44}},\ \bibinfo
  {pages} {943} (\bibinfo {year} {1991})}\BibitemShut {NoStop}%
\bibitem [{\citenamefont {Anisimov}\ \emph {et~al.}(1997)\citenamefont
  {Anisimov}, \citenamefont {Aryasetiawan},\ and\ \citenamefont
  {Lichtenstein}}]{ldau2}%
  \BibitemOpen
  \bibfield  {author} {\bibinfo {author} {\bibfnamefont {V.~I.}\ \bibnamefont
  {Anisimov}}, \bibinfo {author} {\bibfnamefont {F.}~\bibnamefont
  {Aryasetiawan}}, \ and\ \bibinfo {author} {\bibfnamefont {A.~I.}\
  \bibnamefont {Lichtenstein}},\ }\href
  {http://iopscience.iop.org/article/10.1088/0953-8984/9/4/002} {\bibfield
  {journal} {\bibinfo  {journal} {J. Phys: Condens. Matter}\ }\textbf {\bibinfo
  {volume} {9}},\ \bibinfo {pages} {767} (\bibinfo {year} {1997})}\BibitemShut
  {NoStop}%
\bibitem [{\citenamefont {Solovyev}\ \emph {et~al.}(1998)\citenamefont
  {Solovyev}, \citenamefont {Liechtenstein},\ and\ \citenamefont
  {Terakura}}]{solov}%
  \BibitemOpen
  \bibfield  {author} {\bibinfo {author} {\bibfnamefont {I.~V.}\ \bibnamefont
  {Solovyev}}, \bibinfo {author} {\bibfnamefont {A.~I.}\ \bibnamefont
  {Liechtenstein}}, \ and\ \bibinfo {author} {\bibfnamefont {K.}~\bibnamefont
  {Terakura}},\ }\href {\doibase 10.1103/PhysRevLett.80.5758} {\bibfield
  {journal} {\bibinfo  {journal} {Phys. Rev. Lett.}\ }\textbf {\bibinfo
  {volume} {80}},\ \bibinfo {pages} {5758} (\bibinfo {year}
  {1998})}\BibitemShut {NoStop}%
\bibitem [{\citenamefont {Peters}\ \emph {et~al.}(2014)\citenamefont {Peters},
  \citenamefont {Di~Marco}, \citenamefont {Thunstr\"om}, \citenamefont
  {Katsnelson}, \citenamefont {Kirilyuk},\ and\ \citenamefont
  {Eriksson}}]{peters}%
  \BibitemOpen
  \bibfield  {author} {\bibinfo {author} {\bibfnamefont {L.}~\bibnamefont
  {Peters}}, \bibinfo {author} {\bibfnamefont {I.}~\bibnamefont {Di~Marco}},
  \bibinfo {author} {\bibfnamefont {P.}~\bibnamefont {Thunstr\"om}}, \bibinfo
  {author} {\bibfnamefont {M.~I.}\ \bibnamefont {Katsnelson}}, \bibinfo
  {author} {\bibfnamefont {A.}~\bibnamefont {Kirilyuk}}, \ and\ \bibinfo
  {author} {\bibfnamefont {O.}~\bibnamefont {Eriksson}},\ }\href {\doibase
  10.1103/PhysRevB.89.205109} {\bibfield  {journal} {\bibinfo  {journal} {Phys.
  Rev. B}\ }\textbf {\bibinfo {volume} {89}},\ \bibinfo {pages} {205109}
  (\bibinfo {year} {2014})}\BibitemShut {NoStop}%
\bibitem [{\citenamefont {Cococcioni}\ and\ \citenamefont
  {de~Gironcoli}(2005)}]{coco}%
  \BibitemOpen
  \bibfield  {author} {\bibinfo {author} {\bibfnamefont {M.}~\bibnamefont
  {Cococcioni}}\ and\ \bibinfo {author} {\bibfnamefont {S.}~\bibnamefont
  {de~Gironcoli}},\ }\href {\doibase 10.1103/PhysRevB.71.035105} {\bibfield
  {journal} {\bibinfo  {journal} {Phys. Rev. B}\ }\textbf {\bibinfo {volume}
  {71}},\ \bibinfo {pages} {035105} (\bibinfo {year} {2005})}\BibitemShut
  {NoStop}%
\bibitem [{\citenamefont {Bultmark}\ \emph {et~al.}(2009)\citenamefont
  {Bultmark}, \citenamefont {Cricchio}, \citenamefont {Gr\aa{}n\"as},\ and\
  \citenamefont {Nordstr\"om}}]{mult-lars}%
  \BibitemOpen
  \bibfield  {author} {\bibinfo {author} {\bibfnamefont {F.}~\bibnamefont
  {Bultmark}}, \bibinfo {author} {\bibfnamefont {F.}~\bibnamefont {Cricchio}},
  \bibinfo {author} {\bibfnamefont {O.}~\bibnamefont {Gr\aa{}n\"as}}, \ and\
  \bibinfo {author} {\bibfnamefont {L.}~\bibnamefont {Nordstr\"om}},\ }\href
  {\doibase 10.1103/PhysRevB.80.035121} {\bibfield  {journal} {\bibinfo
  {journal} {Phys. Rev. B}\ }\textbf {\bibinfo {volume} {80}},\ \bibinfo
  {pages} {035121} (\bibinfo {year} {2009})}\BibitemShut {NoStop}%
\bibitem [{\citenamefont {Solovyev}\ \emph {et~al.}(1994)\citenamefont
  {Solovyev}, \citenamefont {Dederichs},\ and\ \citenamefont
  {Anisimov}}]{dc-fll}%
  \BibitemOpen
  \bibfield  {author} {\bibinfo {author} {\bibfnamefont {I.~V.}\ \bibnamefont
  {Solovyev}}, \bibinfo {author} {\bibfnamefont {P.~H.}\ \bibnamefont
  {Dederichs}}, \ and\ \bibinfo {author} {\bibfnamefont {V.~I.}\ \bibnamefont
  {Anisimov}},\ }\href {\doibase 10.1103/PhysRevB.50.16861} {\bibfield
  {journal} {\bibinfo  {journal} {Phys. Rev. B}\ }\textbf {\bibinfo {volume}
  {50}},\ \bibinfo {pages} {16861} (\bibinfo {year} {1994})}\BibitemShut
  {NoStop}%
\bibitem [{\citenamefont {Czy\ifmmode~\dot{z}\else \.{z}\fi{}yk}\ and\
  \citenamefont {Sawatzky}(1994)}]{czyzyk1994}%
  \BibitemOpen
  \bibfield  {author} {\bibinfo {author} {\bibfnamefont {M.~T.}\ \bibnamefont
  {Czy\ifmmode~\dot{z}\else \.{z}\fi{}yk}}\ and\ \bibinfo {author}
  {\bibfnamefont {G.~A.}\ \bibnamefont {Sawatzky}},\ }\href {\doibase
  10.1103/PhysRevB.49.14211} {\bibfield  {journal} {\bibinfo  {journal} {Phys.
  Rev. B}\ }\textbf {\bibinfo {volume} {49}},\ \bibinfo {pages} {14211}
  (\bibinfo {year} {1994})}\BibitemShut {NoStop}%
\bibitem [{\citenamefont {Chen}\ and\ \citenamefont {Millis}(2016)}]{lda-lsda}%
  \BibitemOpen
  \bibfield  {author} {\bibinfo {author} {\bibfnamefont {H.}~\bibnamefont
  {Chen}}\ and\ \bibinfo {author} {\bibfnamefont {A.~J.}\ \bibnamefont
  {Millis}},\ }\href {\doibase 10.1103/PhysRevB.93.045133} {\bibfield
  {journal} {\bibinfo  {journal} {Phys. Rev. B}\ }\textbf {\bibinfo {volume}
  {93}},\ \bibinfo {pages} {045133} (\bibinfo {year} {2016})}\BibitemShut
  {NoStop}%
\bibitem [{\citenamefont {Keshavarz}\ \emph {et~al.}(2018)\citenamefont
  {Keshavarz}, \citenamefont {Sch\"ott}, \citenamefont {Millis},\ and\
  \citenamefont {Kvashnin}}]{samara-6}%
  \BibitemOpen
  \bibfield  {author} {\bibinfo {author} {\bibfnamefont {S.}~\bibnamefont
  {Keshavarz}}, \bibinfo {author} {\bibfnamefont {J.}~\bibnamefont {Sch\"ott}},
  \bibinfo {author} {\bibfnamefont {A.~J.}\ \bibnamefont {Millis}}, \ and\
  \bibinfo {author} {\bibfnamefont {Y.~O.}\ \bibnamefont {Kvashnin}},\ }\href
  {\doibase 10.1103/PhysRevB.97.184404} {\bibfield  {journal} {\bibinfo
  {journal} {Phys. Rev. B}\ }\textbf {\bibinfo {volume} {97}},\ \bibinfo
  {pages} {184404} (\bibinfo {year} {2018})}\BibitemShut {NoStop}%
\bibitem [{\citenamefont {Jomard}\ \emph {et~al.}(2008)\citenamefont {Jomard},
  \citenamefont {Amadon}, \citenamefont {Bottin},\ and\ \citenamefont
  {Torrent}}]{amadon-pt}%
  \BibitemOpen
  \bibfield  {author} {\bibinfo {author} {\bibfnamefont {G.}~\bibnamefont
  {Jomard}}, \bibinfo {author} {\bibfnamefont {B.}~\bibnamefont {Amadon}},
  \bibinfo {author} {\bibfnamefont {F.~m.~c.}\ \bibnamefont {Bottin}}, \ and\
  \bibinfo {author} {\bibfnamefont {M.}~\bibnamefont {Torrent}},\ }\href
  {\doibase 10.1103/PhysRevB.78.075125} {\bibfield  {journal} {\bibinfo
  {journal} {Phys. Rev. B}\ }\textbf {\bibinfo {volume} {78}},\ \bibinfo
  {pages} {075125} (\bibinfo {year} {2008})}\BibitemShut {NoStop}%
\bibitem [{\citenamefont {Dorado}\ \emph {et~al.}(2009)\citenamefont {Dorado},
  \citenamefont {Amadon}, \citenamefont {Freyss},\ and\ \citenamefont
  {Bertolus}}]{amadon-uo2}%
  \BibitemOpen
  \bibfield  {author} {\bibinfo {author} {\bibfnamefont {B.}~\bibnamefont
  {Dorado}}, \bibinfo {author} {\bibfnamefont {B.}~\bibnamefont {Amadon}},
  \bibinfo {author} {\bibfnamefont {M.}~\bibnamefont {Freyss}}, \ and\ \bibinfo
  {author} {\bibfnamefont {M.}~\bibnamefont {Bertolus}},\ }\href {\doibase
  10.1103/PhysRevB.79.235125} {\bibfield  {journal} {\bibinfo  {journal} {Phys.
  Rev. B}\ }\textbf {\bibinfo {volume} {79}},\ \bibinfo {pages} {235125}
  (\bibinfo {year} {2009})}\BibitemShut {NoStop}%
\bibitem [{\citenamefont {Zhang}\ \emph {et~al.}(2009)\citenamefont {Zhang},
  \citenamefont {Koepernik}, \citenamefont {Richter},\ and\ \citenamefont
  {Eschrig}}]{koper-coo}%
  \BibitemOpen
  \bibfield  {author} {\bibinfo {author} {\bibfnamefont {W.}~\bibnamefont
  {Zhang}}, \bibinfo {author} {\bibfnamefont {K.}~\bibnamefont {Koepernik}},
  \bibinfo {author} {\bibfnamefont {M.}~\bibnamefont {Richter}}, \ and\
  \bibinfo {author} {\bibfnamefont {H.}~\bibnamefont {Eschrig}},\ }\href
  {\doibase 10.1103/PhysRevB.79.155123} {\bibfield  {journal} {\bibinfo
  {journal} {Phys. Rev. B}\ }\textbf {\bibinfo {volume} {79}},\ \bibinfo
  {pages} {155123} (\bibinfo {year} {2009})}\BibitemShut {NoStop}%
\bibitem [{\citenamefont {Ylvisaker}\ \emph {et~al.}(2009)\citenamefont
  {Ylvisaker}, \citenamefont {Pickett},\ and\ \citenamefont
  {Koepernik}}]{koper-anis}%
  \BibitemOpen
  \bibfield  {author} {\bibinfo {author} {\bibfnamefont {E.~R.}\ \bibnamefont
  {Ylvisaker}}, \bibinfo {author} {\bibfnamefont {W.~E.}\ \bibnamefont
  {Pickett}}, \ and\ \bibinfo {author} {\bibfnamefont {K.}~\bibnamefont
  {Koepernik}},\ }\href {\doibase 10.1103/PhysRevB.79.035103} {\bibfield
  {journal} {\bibinfo  {journal} {Phys. Rev. B}\ }\textbf {\bibinfo {volume}
  {79}},\ \bibinfo {pages} {035103} (\bibinfo {year} {2009})}\BibitemShut
  {NoStop}%
\bibitem [{\citenamefont {Kasinathan}\ \emph {et~al.}(2007)\citenamefont
  {Kasinathan}, \citenamefont {Koepernik}, \citenamefont {Nitzsche},\ and\
  \citenamefont {Rosner}}]{koper-cs}%
  \BibitemOpen
  \bibfield  {author} {\bibinfo {author} {\bibfnamefont {D.}~\bibnamefont
  {Kasinathan}}, \bibinfo {author} {\bibfnamefont {K.}~\bibnamefont
  {Koepernik}}, \bibinfo {author} {\bibfnamefont {U.}~\bibnamefont {Nitzsche}},
  \ and\ \bibinfo {author} {\bibfnamefont {H.}~\bibnamefont {Rosner}},\ }\href
  {\doibase 10.1103/PhysRevLett.99.247210} {\bibfield  {journal} {\bibinfo
  {journal} {Phys. Rev. Lett.}\ }\textbf {\bibinfo {volume} {99}},\ \bibinfo
  {pages} {247210} (\bibinfo {year} {2007})}\BibitemShut {NoStop}%
\bibitem [{\citenamefont {Shick}\ \emph {et~al.}(2001)\citenamefont {Shick},
  \citenamefont {Pickett},\ and\ \citenamefont {Liechtenstein}}]{gamma-ce}%
  \BibitemOpen
  \bibfield  {author} {\bibinfo {author} {\bibfnamefont {A.}~\bibnamefont
  {Shick}}, \bibinfo {author} {\bibfnamefont {W.}~\bibnamefont {Pickett}}, \
  and\ \bibinfo {author} {\bibfnamefont {A.}~\bibnamefont {Liechtenstein}},\
  }\href {\doibase 10.1016/S0368-2048(00)00394-7} {\bibfield  {journal}
  {\bibinfo  {journal} {Journal of Electron Spectroscopy and Related
  Phenomena}\ }\textbf {\bibinfo {volume} {114-116}},\ \bibinfo {pages} {753 }
  (\bibinfo {year} {2001})},\ \bibinfo {note} {proceeding of the Eight
  International Conference on Electronic Spectroscopy and
  Structure,}\BibitemShut {NoStop}%
\bibitem [{\citenamefont {Meredig}\ \emph {et~al.}(2010)\citenamefont
  {Meredig}, \citenamefont {Thompson}, \citenamefont {Hansen}, \citenamefont
  {Wolverton},\ and\ \citenamefont {van~de Walle}}]{u-ramping}%
  \BibitemOpen
  \bibfield  {author} {\bibinfo {author} {\bibfnamefont {B.}~\bibnamefont
  {Meredig}}, \bibinfo {author} {\bibfnamefont {A.}~\bibnamefont {Thompson}},
  \bibinfo {author} {\bibfnamefont {H.~A.}\ \bibnamefont {Hansen}}, \bibinfo
  {author} {\bibfnamefont {C.}~\bibnamefont {Wolverton}}, \ and\ \bibinfo
  {author} {\bibfnamefont {A.}~\bibnamefont {van~de Walle}},\ }\href {\doibase
  10.1103/PhysRevB.82.195128} {\bibfield  {journal} {\bibinfo  {journal} {Phys.
  Rev. B}\ }\textbf {\bibinfo {volume} {82}},\ \bibinfo {pages} {195128}
  (\bibinfo {year} {2010})}\BibitemShut {NoStop}%
\bibitem [{\citenamefont {Gryaznov}\ \emph {et~al.}(2012)\citenamefont
  {Gryaznov}, \citenamefont {Heifets},\ and\ \citenamefont
  {Kotomin}}]{cont-sym-red}%
  \BibitemOpen
  \bibfield  {author} {\bibinfo {author} {\bibfnamefont {D.}~\bibnamefont
  {Gryaznov}}, \bibinfo {author} {\bibfnamefont {E.}~\bibnamefont {Heifets}}, \
  and\ \bibinfo {author} {\bibfnamefont {E.}~\bibnamefont {Kotomin}},\ }\href
  {\doibase 10.1039/C2CP40297A} {\bibfield  {journal} {\bibinfo  {journal}
  {Phys. Chem. Chem. Phys.}\ }\textbf {\bibinfo {volume} {14}},\ \bibinfo
  {pages} {4482} (\bibinfo {year} {2012})}\BibitemShut {NoStop}%
\bibitem [{\citenamefont {Geng}\ \emph {et~al.}(2010)\citenamefont {Geng},
  \citenamefont {Chen}, \citenamefont {Kaneta}, \citenamefont {Kinoshita},\
  and\ \citenamefont {Wu}}]{kinetic}%
  \BibitemOpen
  \bibfield  {author} {\bibinfo {author} {\bibfnamefont {H.~Y.}\ \bibnamefont
  {Geng}}, \bibinfo {author} {\bibfnamefont {Y.}~\bibnamefont {Chen}}, \bibinfo
  {author} {\bibfnamefont {Y.}~\bibnamefont {Kaneta}}, \bibinfo {author}
  {\bibfnamefont {M.}~\bibnamefont {Kinoshita}}, \ and\ \bibinfo {author}
  {\bibfnamefont {Q.}~\bibnamefont {Wu}},\ }\href {\doibase
  10.1103/PhysRevB.82.094106} {\bibfield  {journal} {\bibinfo  {journal} {Phys.
  Rev. B}\ }\textbf {\bibinfo {volume} {82}},\ \bibinfo {pages} {094106}
  (\bibinfo {year} {2010})}\BibitemShut {NoStop}%
\bibitem [{rsp()}]{rspt-web}%
  \BibitemOpen
  \href@noop {} {\enquote {\bibinfo {title} {{RSP}t - relativistic spin
  polarized toolkit},}\ }\bibinfo {howpublished}
  {\url{http://fplmto-rspt.org/}}\BibitemShut {NoStop}%
\bibitem [{\citenamefont {Wills}\ \emph {et~al.}(2010)\citenamefont {Wills},
  \citenamefont {Alouani}, \citenamefont {Andersson}, \citenamefont {Delin},
  \citenamefont {Eriksson},\ and\ \citenamefont {Grechnyev}}]{rspt-book}%
  \BibitemOpen
  \bibfield  {author} {\bibinfo {author} {\bibfnamefont {J.~M.}\ \bibnamefont
  {Wills}}, \bibinfo {author} {\bibfnamefont {M.}~\bibnamefont {Alouani}},
  \bibinfo {author} {\bibfnamefont {P.}~\bibnamefont {Andersson}}, \bibinfo
  {author} {\bibfnamefont {A.}~\bibnamefont {Delin}}, \bibinfo {author}
  {\bibfnamefont {O.}~\bibnamefont {Eriksson}}, \ and\ \bibinfo {author}
  {\bibfnamefont {O.}~\bibnamefont {Grechnyev}},\ }\href@noop {} {\emph
  {\bibinfo {title} {Full-Potential Electronic Structure Method}}},\ edited by\
  \bibinfo {editor} {\bibfnamefont {E.~S.}\ \bibnamefont {H.~Dreysse}}\ and\
  \bibinfo {editor} {\bibfnamefont {P.~P.}\ \bibnamefont {of~Solids: Springer
  Series~in Solid-State~Sciences}}\ (\bibinfo  {publisher} {Springer-Verlag},\
  \bibinfo {address} {Berlin},\ \bibinfo {year} {2010})\BibitemShut {NoStop}%
\bibitem [{\citenamefont {Grechnev}\ \emph {et~al.}(2007)\citenamefont
  {Grechnev}, \citenamefont {Di~Marco}, \citenamefont {Katsnelson},
  \citenamefont {Lichtenstein}, \citenamefont {Wills},\ and\ \citenamefont
  {Eriksson}}]{rspt-lda+u}%
  \BibitemOpen
  \bibfield  {author} {\bibinfo {author} {\bibfnamefont {A.}~\bibnamefont
  {Grechnev}}, \bibinfo {author} {\bibfnamefont {I.}~\bibnamefont {Di~Marco}},
  \bibinfo {author} {\bibfnamefont {M.~I.}\ \bibnamefont {Katsnelson}},
  \bibinfo {author} {\bibfnamefont {A.~I.}\ \bibnamefont {Lichtenstein}},
  \bibinfo {author} {\bibfnamefont {J.}~\bibnamefont {Wills}}, \ and\ \bibinfo
  {author} {\bibfnamefont {O.}~\bibnamefont {Eriksson}},\ }\href {\doibase
  10.1103/PhysRevB.76.035107} {\bibfield  {journal} {\bibinfo  {journal} {Phys.
  Rev. B}\ }\textbf {\bibinfo {volume} {76}},\ \bibinfo {pages} {035107}
  (\bibinfo {year} {2007})}\BibitemShut {NoStop}%
\bibitem [{\citenamefont {Gr\aa{}n\"as}\ \emph {et~al.}(2012)\citenamefont
  {Gr\aa{}n\"as}, \citenamefont {Marco}, \citenamefont {Thunstr\"om},
  \citenamefont {Nordstr\"om}, \citenamefont {Eriksson}, \citenamefont
  {Bj\"orkman},\ and\ \citenamefont {Wills}}]{rspt-oscar}%
  \BibitemOpen
  \bibfield  {author} {\bibinfo {author} {\bibfnamefont {O.}~\bibnamefont
  {Gr\aa{}n\"as}}, \bibinfo {author} {\bibfnamefont {I.~D.}\ \bibnamefont
  {Marco}}, \bibinfo {author} {\bibfnamefont {P.}~\bibnamefont {Thunstr\"om}},
  \bibinfo {author} {\bibfnamefont {L.}~\bibnamefont {Nordstr\"om}}, \bibinfo
  {author} {\bibfnamefont {O.}~\bibnamefont {Eriksson}}, \bibinfo {author}
  {\bibfnamefont {T.}~\bibnamefont {Bj\"orkman}}, \ and\ \bibinfo {author}
  {\bibfnamefont {J.}~\bibnamefont {Wills}},\ }\href {\doibase
  10.1016/j.commatsci.2011.11.032} {\bibfield  {journal} {\bibinfo  {journal}
  {Computational Materials Science}\ }\textbf {\bibinfo {volume} {55}},\
  \bibinfo {pages} {295 } (\bibinfo {year} {2012})}\BibitemShut {NoStop}%
\bibitem [{\citenamefont {Shick}\ \emph {et~al.}(2005)\citenamefont {Shick},
  \citenamefont {Drchal},\ and\ \citenamefont {Havela}}]{4-index-u}%
  \BibitemOpen
  \bibfield  {author} {\bibinfo {author} {\bibfnamefont {A.~B.}\ \bibnamefont
  {Shick}}, \bibinfo {author} {\bibfnamefont {V.}~\bibnamefont {Drchal}}, \
  and\ \bibinfo {author} {\bibfnamefont {L.}~\bibnamefont {Havela}},\ }\href
  {http://stacks.iop.org/0295-5075/69/i=4/a=588} {\bibfield  {journal}
  {\bibinfo  {journal} {EPL (Europhysics Letters)}\ }\textbf {\bibinfo {volume}
  {69}},\ \bibinfo {pages} {588} (\bibinfo {year} {2005})}\BibitemShut
  {NoStop}%
\bibitem [{\citenamefont {Fang}\ \emph {et~al.}(1999)\citenamefont {Fang},
  \citenamefont {Solovyev}, \citenamefont {Sawada},\ and\ \citenamefont
  {Terakura}}]{u-tmo1}%
  \BibitemOpen
  \bibfield  {author} {\bibinfo {author} {\bibfnamefont {Z.}~\bibnamefont
  {Fang}}, \bibinfo {author} {\bibfnamefont {I.~V.}\ \bibnamefont {Solovyev}},
  \bibinfo {author} {\bibfnamefont {H.}~\bibnamefont {Sawada}}, \ and\ \bibinfo
  {author} {\bibfnamefont {K.}~\bibnamefont {Terakura}},\ }\href {\doibase
  10.1103/PhysRevB.59.762} {\bibfield  {journal} {\bibinfo  {journal} {Phys.
  Rev. B}\ }\textbf {\bibinfo {volume} {59}},\ \bibinfo {pages} {762} (\bibinfo
  {year} {1999})}\BibitemShut {NoStop}%
\bibitem [{\citenamefont {Dudarev}\ \emph {et~al.}(1998)\citenamefont
  {Dudarev}, \citenamefont {Botton}, \citenamefont {Savrasov}, \citenamefont
  {Humphreys},\ and\ \citenamefont {Sutton}}]{u-tmo2}%
  \BibitemOpen
  \bibfield  {author} {\bibinfo {author} {\bibfnamefont {S.~L.}\ \bibnamefont
  {Dudarev}}, \bibinfo {author} {\bibfnamefont {G.~A.}\ \bibnamefont {Botton}},
  \bibinfo {author} {\bibfnamefont {S.~Y.}\ \bibnamefont {Savrasov}}, \bibinfo
  {author} {\bibfnamefont {C.~J.}\ \bibnamefont {Humphreys}}, \ and\ \bibinfo
  {author} {\bibfnamefont {A.~P.}\ \bibnamefont {Sutton}},\ }\href {\doibase
  10.1103/PhysRevB.57.1505} {\bibfield  {journal} {\bibinfo  {journal} {Phys.
  Rev. B}\ }\textbf {\bibinfo {volume} {57}},\ \bibinfo {pages} {1505}
  (\bibinfo {year} {1998})}\BibitemShut {NoStop}%
\bibitem [{\citenamefont {Kotani}\ and\ \citenamefont
  {Yamazaki}(1992)}]{u-j-uo2}%
  \BibitemOpen
  \bibfield  {author} {\bibinfo {author} {\bibfnamefont {A.}~\bibnamefont
  {Kotani}}\ and\ \bibinfo {author} {\bibfnamefont {T.}~\bibnamefont
  {Yamazaki}},\ }\href {\doibase 10.1143/PTPS.108.117} {\bibfield  {journal}
  {\bibinfo  {journal} {Prog. Theor. Phys. Supplement}\ }\textbf {\bibinfo
  {volume} {108}},\ \bibinfo {pages} {117} (\bibinfo {year}
  {1992})}\BibitemShut {NoStop}%
\bibitem [{\citenamefont {Roth}(1958)}]{feo1}%
  \BibitemOpen
  \bibfield  {author} {\bibinfo {author} {\bibfnamefont {W.~L.}\ \bibnamefont
  {Roth}},\ }\href {\doibase 10.1103/PhysRev.110.1333} {\bibfield  {journal}
  {\bibinfo  {journal} {Phys. Rev.}\ }\textbf {\bibinfo {volume} {110}},\
  \bibinfo {pages} {1333} (\bibinfo {year} {1958})}\BibitemShut {NoStop}%
\bibitem [{\citenamefont {Stafford}(2006)}]{randfixedsum}%
  \BibitemOpen
  \bibfield  {author} {\bibinfo {author} {\bibfnamefont {R.}~\bibnamefont
  {Stafford}},\ }\href
  {https://se.mathworks.com/matlabcentral/fileexchange/9700-random-vectors-with-fixed-sum}
  {\enquote {\bibinfo {title} {Random vectors with fixed sum},}\ } (\bibinfo
  {year} {2006})\BibitemShut {NoStop}%
\bibitem [{Note1()}]{Note1}%
  \BibitemOpen
  \bibinfo {note} {In the present work we sampled the interior in a slightly
  non-uniform way. If a spin channel is less than half-filled, $N_{\sigma } <
  M/2$ where $M$ is the number of orbitals, we simply set the occupation of the
  corresponding spin-orbitals to a random number uniformly distributed between
  0 and 1, and after that rescale all the elements to obtain the given filling
  $N_{\sigma }$. The scaling bias the sampling away from the already covered
  corner cases. If the spin channel is set to be more than half-filled
  ($N_{\sigma } > M/2$) we restrict the initial distribution to be between
  $N_{\sigma }/M$ and 1, where M is the number of orbitals, before the
  rescaling of the elements. This restriction gradually excludes points close
  to the corners, {\protect \em i.e.} simulations with almost empty orbitals,
  as the filling of the spin-channel becomes larger. All matrices $D$ with
  elements that become larger than 1 after the rescaling were
  discarded.}\BibitemShut {Stop}%
\bibitem [{\citenamefont {Hugel}\ and\ \citenamefont {Kamal}(1996)}]{kamal}%
  \BibitemOpen
  \bibfield  {author} {\bibinfo {author} {\bibfnamefont {J.}~\bibnamefont
  {Hugel}}\ and\ \bibinfo {author} {\bibfnamefont {M.}~\bibnamefont {Kamal}},\
  }\href {\doibase 10.1016/0038-1098(96)00477-2} {\bibfield  {journal}
  {\bibinfo  {journal} {Solid State Communications}\ }\textbf {\bibinfo
  {volume} {100}},\ \bibinfo {pages} {457 } (\bibinfo {year}
  {1996})}\BibitemShut {NoStop}%
\bibitem [{\citenamefont {Alperin}(1961)}]{alperin1961}%
  \BibitemOpen
  \bibfield  {author} {\bibinfo {author} {\bibfnamefont {H.~A.}\ \bibnamefont
  {Alperin}},\ }\href {\doibase 10.1103/PhysRevLett.6.55} {\bibfield  {journal}
  {\bibinfo  {journal} {Phys. Rev. Lett.}\ }\textbf {\bibinfo {volume} {6}},\
  \bibinfo {pages} {55} (\bibinfo {year} {1961})}\BibitemShut {NoStop}%
\bibitem [{\citenamefont {Fernandez}\ \emph {et~al.}(1998)\citenamefont
  {Fernandez}, \citenamefont {Vettier}, \citenamefont {de~Bergevin},
  \citenamefont {Giles},\ and\ \citenamefont {Neubeck}}]{fernandez1998}%
  \BibitemOpen
  \bibfield  {author} {\bibinfo {author} {\bibfnamefont {V.}~\bibnamefont
  {Fernandez}}, \bibinfo {author} {\bibfnamefont {C.}~\bibnamefont {Vettier}},
  \bibinfo {author} {\bibfnamefont {F.}~\bibnamefont {de~Bergevin}}, \bibinfo
  {author} {\bibfnamefont {C.}~\bibnamefont {Giles}}, \ and\ \bibinfo {author}
  {\bibfnamefont {W.}~\bibnamefont {Neubeck}},\ }\href {\doibase
  10.1103/PhysRevB.57.7870} {\bibfield  {journal} {\bibinfo  {journal} {Phys.
  Rev. B}\ }\textbf {\bibinfo {volume} {57}},\ \bibinfo {pages} {7870}
  (\bibinfo {year} {1998})}\BibitemShut {NoStop}%
\bibitem [{\citenamefont {Rinaldi-Montes}\ \emph {et~al.}(2016)\citenamefont
  {Rinaldi-Montes}, \citenamefont {Gorria}, \citenamefont {Martínez-Blanco},
  \citenamefont {Fuertes}, \citenamefont {Puente-Orench}, \citenamefont
  {Olivi},\ and\ \citenamefont {Blanco}}]{rinaldi2016}%
  \BibitemOpen
  \bibfield  {author} {\bibinfo {author} {\bibfnamefont {N.}~\bibnamefont
  {Rinaldi-Montes}}, \bibinfo {author} {\bibfnamefont {P.}~\bibnamefont
  {Gorria}}, \bibinfo {author} {\bibfnamefont {D.}~\bibnamefont
  {Martínez-Blanco}}, \bibinfo {author} {\bibfnamefont {A.~B.}\ \bibnamefont
  {Fuertes}}, \bibinfo {author} {\bibfnamefont {I.}~\bibnamefont
  {Puente-Orench}}, \bibinfo {author} {\bibfnamefont {L.}~\bibnamefont
  {Olivi}}, \ and\ \bibinfo {author} {\bibfnamefont {J.~A.}\ \bibnamefont
  {Blanco}},\ }\href {\doibase 10.1063/1.4943062} {\bibfield  {journal}
  {\bibinfo  {journal} {AIP Advances}\ }\textbf {\bibinfo {volume} {6}},\
  \bibinfo {pages} {056104} (\bibinfo {year} {2016})},\ \Eprint
  {http://arxiv.org/abs/https://doi.org/10.1063/1.4943062}
  {https://doi.org/10.1063/1.4943062} \BibitemShut {NoStop}%
\bibitem [{\citenamefont {Brok}\ \emph {et~al.}(2015)\citenamefont {Brok},
  \citenamefont {Lefmann}, \citenamefont {Deen}, \citenamefont {Lebech},
  \citenamefont {Jacobsen}, \citenamefont {Nilsen}, \citenamefont {Keller},\
  and\ \citenamefont {Frandsen}}]{brok2015}%
  \BibitemOpen
  \bibfield  {author} {\bibinfo {author} {\bibfnamefont {E.}~\bibnamefont
  {Brok}}, \bibinfo {author} {\bibfnamefont {K.}~\bibnamefont {Lefmann}},
  \bibinfo {author} {\bibfnamefont {P.~P.}\ \bibnamefont {Deen}}, \bibinfo
  {author} {\bibfnamefont {B.}~\bibnamefont {Lebech}}, \bibinfo {author}
  {\bibfnamefont {H.}~\bibnamefont {Jacobsen}}, \bibinfo {author}
  {\bibfnamefont {G.~J.}\ \bibnamefont {Nilsen}}, \bibinfo {author}
  {\bibfnamefont {L.}~\bibnamefont {Keller}}, \ and\ \bibinfo {author}
  {\bibfnamefont {C.}~\bibnamefont {Frandsen}},\ }\href {\doibase
  10.1103/PhysRevB.91.014431} {\bibfield  {journal} {\bibinfo  {journal} {Phys.
  Rev. B}\ }\textbf {\bibinfo {volume} {91}},\ \bibinfo {pages} {014431}
  (\bibinfo {year} {2015})}\BibitemShut {NoStop}%
\bibitem [{\citenamefont {Schr\"on}\ \emph {et~al.}(2012)\citenamefont
  {Schr\"on}, \citenamefont {R\"odl},\ and\ \citenamefont
  {Bechstedt}}]{schron}%
  \BibitemOpen
  \bibfield  {author} {\bibinfo {author} {\bibfnamefont {A.}~\bibnamefont
  {Schr\"on}}, \bibinfo {author} {\bibfnamefont {C.}~\bibnamefont {R\"odl}}, \
  and\ \bibinfo {author} {\bibfnamefont {F.}~\bibnamefont {Bechstedt}},\ }\href
  {\doibase 10.1103/PhysRevB.86.115134} {\bibfield  {journal} {\bibinfo
  {journal} {Phys. Rev. B}\ }\textbf {\bibinfo {volume} {86}},\ \bibinfo
  {pages} {115134} (\bibinfo {year} {2012})}\BibitemShut {NoStop}%
\bibitem [{\citenamefont {Herrmann-Ronzaud}\ \emph {et~al.}(1978)\citenamefont
  {Herrmann-Ronzaud}, \citenamefont {Burlet},\ and\ \citenamefont
  {Rossat-Mignod}}]{feo2}%
  \BibitemOpen
  \bibfield  {author} {\bibinfo {author} {\bibfnamefont {D.}~\bibnamefont
  {Herrmann-Ronzaud}}, \bibinfo {author} {\bibfnamefont {P.}~\bibnamefont
  {Burlet}}, \ and\ \bibinfo {author} {\bibfnamefont {J.}~\bibnamefont
  {Rossat-Mignod}},\ }\href {http://stacks.iop.org/0022-3719/11/i=10/a=023}
  {\bibfield  {journal} {\bibinfo  {journal} {J. Phys. C}\ }\textbf {\bibinfo
  {volume} {11}},\ \bibinfo {pages} {2123} (\bibinfo {year}
  {1978})}\BibitemShut {NoStop}%
\bibitem [{\citenamefont {Jauch}\ \emph {et~al.}(2001)\citenamefont {Jauch},
  \citenamefont {Reehuis}, \citenamefont {Bleif}, \citenamefont {Kubanek},\
  and\ \citenamefont {Pattison}}]{coo-exp1}%
  \BibitemOpen
  \bibfield  {author} {\bibinfo {author} {\bibfnamefont {W.}~\bibnamefont
  {Jauch}}, \bibinfo {author} {\bibfnamefont {M.}~\bibnamefont {Reehuis}},
  \bibinfo {author} {\bibfnamefont {H.~J.}\ \bibnamefont {Bleif}}, \bibinfo
  {author} {\bibfnamefont {F.}~\bibnamefont {Kubanek}}, \ and\ \bibinfo
  {author} {\bibfnamefont {P.}~\bibnamefont {Pattison}},\ }\href {\doibase
  10.1103/PhysRevB.64.052102} {\bibfield  {journal} {\bibinfo  {journal} {Phys.
  Rev. B}\ }\textbf {\bibinfo {volume} {64}},\ \bibinfo {pages} {052102}
  (\bibinfo {year} {2001})}\BibitemShut {NoStop}%
\bibitem [{\citenamefont {Fjellv\aa{}g}\ \emph {et~al.}(1996)\citenamefont
  {Fjellv\aa{}g}, \citenamefont {Gr\o{}nvold}, \citenamefont {St\o{}len},\ and\
  \citenamefont {Hauback}}]{feo-exp1}%
  \BibitemOpen
  \bibfield  {author} {\bibinfo {author} {\bibfnamefont {H.}~\bibnamefont
  {Fjellv\aa{}g}}, \bibinfo {author} {\bibfnamefont {F.}~\bibnamefont
  {Gr\o{}nvold}}, \bibinfo {author} {\bibfnamefont {S.}~\bibnamefont
  {St\o{}len}}, \ and\ \bibinfo {author} {\bibfnamefont {B.}~\bibnamefont
  {Hauback}},\ }\href {\doibase 10.1006/jssc.1996.0206} {\bibfield  {journal}
  {\bibinfo  {journal} {J. Solid State Chemistry}\ }\textbf {\bibinfo {volume}
  {124}},\ \bibinfo {pages} {52 } (\bibinfo {year} {1996})}\BibitemShut
  {NoStop}%
\bibitem [{\citenamefont {Fjellv\aa{}g}\ \emph {et~al.}(2002)\citenamefont
  {Fjellv\aa{}g}, \citenamefont {Hauback}, \citenamefont {Vogt},\ and\
  \citenamefont {St\o{}len}}]{feo-exp2}%
  \BibitemOpen
  \bibfield  {author} {\bibinfo {author} {\bibfnamefont {H.}~\bibnamefont
  {Fjellv\aa{}g}}, \bibinfo {author} {\bibfnamefont {B.~C.}\ \bibnamefont
  {Hauback}}, \bibinfo {author} {\bibfnamefont {T.}~\bibnamefont {Vogt}}, \
  and\ \bibinfo {author} {\bibfnamefont {S.}~\bibnamefont {St\o{}len}},\ }\href
  {http://ammin.geoscienceworld.org/content/87/2-3/347.abstract} {\bibfield
  {journal} {\bibinfo  {journal} {American Mineralogist}\ }\textbf {\bibinfo
  {volume} {87}},\ \bibinfo {pages} {347} (\bibinfo {year} {2002})}\BibitemShut
  {NoStop}%
\bibitem [{\citenamefont {Schr\"on}\ and\ \citenamefont
  {Bechstedt}(2013)}]{schron-mom}%
  \BibitemOpen
  \bibfield  {author} {\bibinfo {author} {\bibfnamefont {A.}~\bibnamefont
  {Schr\"on}}\ and\ \bibinfo {author} {\bibfnamefont {F.}~\bibnamefont
  {Bechstedt}},\ }\href {http://stacks.iop.org/0953-8984/25/i=48/a=486002}
  {\bibfield  {journal} {\bibinfo  {journal} {J. Phys: Condens. Matter}\
  }\textbf {\bibinfo {volume} {25}},\ \bibinfo {pages} {486002} (\bibinfo
  {year} {2013})}\BibitemShut {NoStop}%
\end{thebibliography}
\bibliographystyle{apsrev4-1}

\end{document}